\documentclass[twocolumn,prb,amsmath,amssymb,floatfix,superscriptaddress]{revtex4-2} 
\usepackage{amsmath}
\usepackage{amssymb}
\usepackage{graphicx}
\usepackage{times}
\usepackage{bm}
\usepackage{color}
\usepackage{hyperref}
\usepackage{graphicx}
\usepackage{upgreek}
\usepackage{scalerel}
\usepackage{multirow}
 
\usepackage{pgffor}
\usepackage{pdfpages}
\usepackage{mathdots}
\usepackage{float}
\usepackage{silence}
\usepackage{physics} 
\usepackage{lscape}   
\usepackage{color, colortbl}
\usepackage[table]{xcolor}
\usepackage{comment}  
\makeatletter
\AtBeginDocument{\let\LS@rot\@undefined}
\makeatother

\newcommand{\nn}{\nonumber}
\newcommand{\beq}{\begin{equation}}
\newcommand{\eeq}{\end{equation}}

\begin{document}

\title{Detecting the Topological Phase Transition in Superconductor-Semiconductor Hybrids by \\ 
Electronic Raman Spectroscopy}

\author{Takeshi Mizushima}
\affiliation{Department of Materials Engineering Science, Osaka University, Toyonaka, Osaka 560-8531, Japan}

\author{Yukio Tanaka}
\affiliation{Department of Applied Physics, Nagoya University, Nagoya 464-8603, Japan}
\affiliation{Research Center for Crystalline Materials Engineering, Nagoya University, Nagoya 464-8603, Japan}

\author{Jorge Cayao}
\affiliation{Department of Physics and Astronomy, Uppsala University, Box 516, S-751 20 Uppsala, Sweden}

\date{\today}
\begin{abstract}
In superconductor-semiconductor hybrids, applying a magnetic field closes a trivial bulk gap and causes a topological phase transition (TPT), resulting in the emergence of Majorana zero modes at both ends of the wires. However, trivial Andreev bound states formed at the interface with metallic leads mimic the local Majorana properties, making it difficult to detect the TPT through local conductance measurements. In this work, we investigate the detection of the TPT by exploiting the static and dynamical density response of the hybrid system. In particular, we demonstrate that the dynamical renormalized responses, the density response including the effect of Coulomb interactions, reveal the characteristic electronic structure and detect the TPT, which we then show to produce strong intensities of Raman scattering. Furthermore, we find that gapless plasmons emerge in the normal state, signaling the bulk Lifshitz transition. Our results thus predict that the bulk response of superconducting nanowires is a powerful spectroscopic approach to detect the bulk topological phase transition.
\end{abstract}
\maketitle

\section{Introduction}

Topological superconductivity has motivated an impressive amount of studies in the past fifteen years due to its possible use for fault tolerant quantum computation \cite{tanaka2011symmetry,ali12,sta13,sarma2015majorana,sat16,sato2017topological,lutchyn2018majorana,zhang2019next,frolov2019quest,org20,aguado2020majorana,Marra_2022,tanaka2024theory}. To realize topological superconductivity, superconductor-semiconductor hybrids attracted most of the attention \cite{sat09,sat10,PhysRevLett.105.077001,PhysRevLett.105.177002,tam21,asa13,esl24}. Applying a magnetic field in the hybrid systems causes the bulk energy gap to close and reopen, involving a change in a topological invariant defined over the entire momentum space. {This gap closing and reopening implies that} electrons in the hybrid system undergo a topological phase transition (TPT) across a critical field. In the topological phase above the critical field, Majorana zero modes (MZMs), self-conjugated zero-energy quasiparticles obeying non-Abelian exchange statistics, emerge at both ends of the wire and define a nonlocal fermion. While braiding MZMs allows for implementing fault-tolerant qubit operations~\cite{nay08}, the unambiguous identification of MZMs and topological superconductivity is still a challenging task~\cite{prada2019andreev}.

There are two ways to identify topological superconductivity: detecting MZMs or measuring the TPT. Regarding the former, despite numerous experiments in superconductor-semiconductor hybrids, unequivocal evidence has yet to be established~\cite{lutchyn2018majorana,zhang2019next,prada2019andreev,tanaka2024theory}. This is because hybrid systems exhibit unavoidable effects, such as disorder~\cite{Pikulin2012A,PhysRevB.107.184519}, nonuniform potentials \cite{kel12,cay15}, and formation of quantum dots~\cite{PhysRevLett.123.117001,PhysRevB.98.245407}, that generically cause the appearance of trivial Andreev bound states~\cite{prada2019andreev}. These states mimic the local properties of MZMs, leading to a nearly quantized local conductance even in the topologically trivial phase~\cite{kel12,pra12,roy13,sta14,cay15,liu17,moo18,PhysRevB.104.134507,sah23}. This makes it difficult to distinguish between MZMs and trivial Andreev bound states through local conductance measurements~\cite{prada2019andreev}, thereby blurring the signal of the TPT.

Besides local measurements, it has been predicted that the bulk gap closing at the TPT can be observed through nonlocal conductance in three-terminal devices~\cite{PhysRevB.97.045421,PhysRevLett.124.036801}. In this regard, a topological gap protocol was proposed in Ref.\,\cite{pik21}, which involves a sequence of local and nonlocal transport measurements and offers an algorithm for automatically identifying regions where MZMs exist~\cite{pik21}. Recently, Microsoft Quantum reported that state-of-the-art superconductor-semiconductor devices passed the topological gap protocol, suggesting a strong likelihood of identifying a topological phase with MZMs~\cite{microsoft}. However, it has been discussed that a trivial Andreev band formed in a semiconducting wire also mimics both nearly quantized local conductance and bulk reopening signatures in nonlocal measurements~\cite{hes23,sar24,hes24,ant23,hen23,akh22,fro23}, which challenges the topological gap protocol. More recently, the Microsoft team has also reported a potential way to identify MZMs using interferometry measurements~\cite{microsoft2025interferometric}, which, however, can neither rule out trivial Andreev bound states nor detect the TPT. Thus, detecting the TPT, crucial for identifying topological superconductivity in superconductor-semiconductor hybrids, is still an open problem.

In this paper, we explore an alternative way for directly measuring the TPT through bulk measurements in experimentally feasible superconductor-semiconductor hybrids. We first demonstrate that the field evolution of the dynamical density-density response function, which includes the Coulomb interaction effect, captures the bulk gap closing and reopening signal across the TPT. Notably, we discover that the dynamical response produces strong spectroscopic intensities that can be detected via electronic Raman spectroscopy \cite{klein84,dev07}, used before for exploring charge density waves and Higgs modes in $2H$-NbSe$_2$~\cite{wang74,tsang76,soor80,soor81,pereira82,wu08,mia11,measson14,varma82,browne83,cea14}, and low-lying excitations in semiconducting wires~\cite{wei89,ege90,gon91,gon93,sch94,dah95,bie96,ste96,sch96,ulr97,boo03,mei21}. As a byproduct, we also find that the field evolution of the response signal measures the bulk Lifshitz transition, which involves a change in the topology of a Fermi surface, and reveals gapless plasmons in the normal state of the semiconducting wire. More precisely, we obtain that one of the gapless plasmons softens as the applied magnetic field increases and vanishes at the Lifshitz transition. Our work puts forward the dynamical response of superconducting nanowires as a robust spectroscopic tool for identifying smoking-gun signatures of the bulk topological phase transition.

\section{Topological phase transition in superconductor-semiconductor hybrids}

\begin{figure}[t!]
\includegraphics[width=85mm]{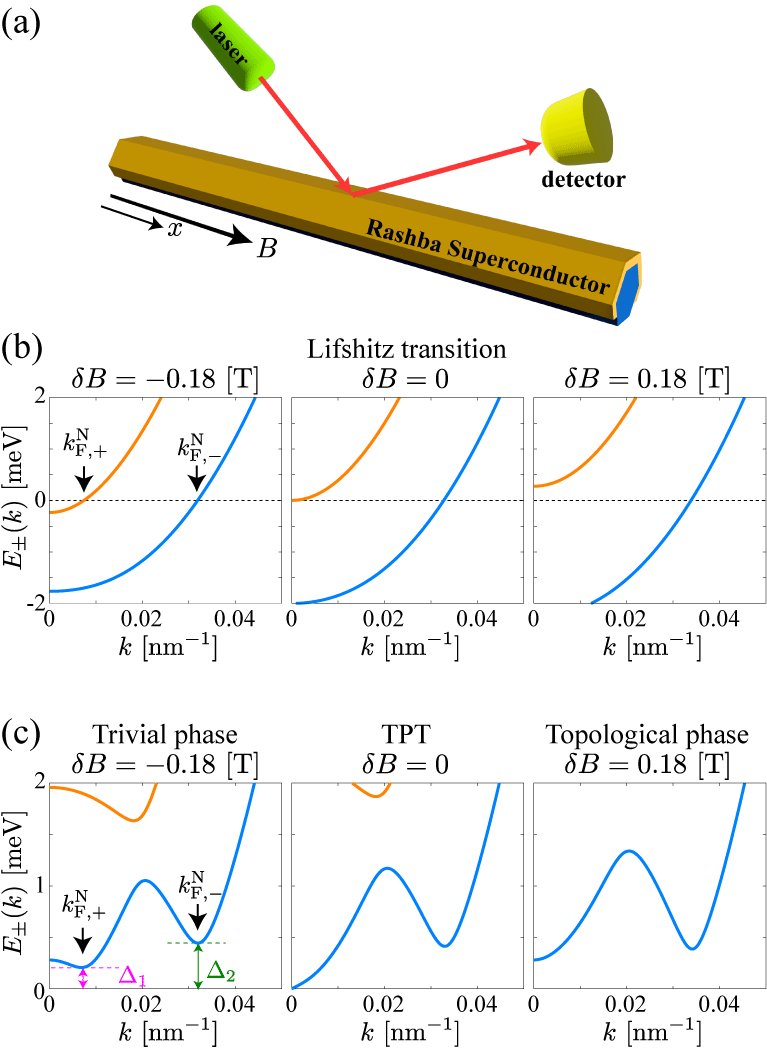}
\caption{(a) Schematics of our setup: Superconducting wire with Rashba spin-orbit coupling, called the Rashba superconductor, under a Zeeman field $B$. Incident light (red) from a laser (green) reaches the Rashba superconductor, where it is scattered and then reaches the detector (yellow). (b) Energy dispersion in the normal state at $\delta B=-0.18~{\rm T}$, the Lifshitz transition $\delta B = 0$ ($B=B_{\rm L}$), and $\delta B=0.18~{\rm T}$, where we set $\Delta = 0$. (c) Energy dispersion in the superconducting state at $\delta B=-0.18~{\rm T}$, the TPT $\delta B = 0$ ($B=B_{\rm c}$), and $\delta B=0.18~{\rm T}$, where $k^{\rm N}_{{\rm F},+}$ ($k^{\rm N}_{{\rm F},-}$) denotes the Fermi wavenumber of the inner (outer) band in the normal state and $\Delta_{1}$ ($\Delta_{2}$) represents the excitation gap at $k=k^{+}_{\rm F}$ ($k=k^{-}_{\rm F}$).}
\label{fig:setup0}
\end{figure}

We consider a single-channel one-dimensional semiconducting wire with Rashba spin-orbit coupling, where the radius of the wire is smaller than the Fermi wavelength. The semiconducting wire is coupled to a conventional $s$-wave superconductor as shown in Fig.~\ref{fig:setup0}(a), enabling superconducting correlation to penetrate into the semiconducting wire by proximity effect. As a result, electrons in the wire experience an effective superconducting gap, here denoted by $\Delta$ and the effective Hamiltonian for electrons in the wire is then described by~\cite{sat09,PhysRevLett.105.077001,PhysRevLett.105.177002,sau10,ali12} 
\begin{align}
\mathcal{H}(k) 
= \xi_k\tau_z + \alpha k \sigma_z \tau_0 + b\sigma_x \tau_z -\Delta\sigma_y\tau_y, 
\label{eq:H}
\end{align}
where $\xi_k\equiv k^2/2m_{\rm eff}-\mu$ is the energy of free electrons with effective mass $m_{\rm eff}$ and chemical potential $\mu$. We set $\hbar =1$ and introduce the $i$-th Pauli matrix in the spin (particle-hole) space as $\sigma_{i}$ ($\tau_{i}$). The Zeeman energy due to the applied magnetic field $B$ is $b\equiv \frac{1}{2}g\mu_{\rm B}B$ with the effective $g$-factor ($g$) and the Bohr magneton ($\mu_{\rm B}$). The second and fourth terms in Eq.~\eqref{eq:H} denote the Rashba spin-orbit coupling and the induced pair potential, respectively. In experiments, semiconductors including heavy elements, such as InAs and InSb, have served as the key component for Majorana devices \cite{lut18,flensberg2021engineered,frolov2019quest} due to their strong Rashba spin-orbit coupling and large Land\'{e} $g$-factor \cite{lut18}. In this work, we take the parameters of the semiconductor InSb, i.e., $m_{\rm eff}=0.014m_{\rm e}$, $\alpha = 0.5~{\rm eV}$\AA, $g=50$, and $\mu = 1.0~{\rm meV}$~\cite{lut18}, where $m_{\rm e}$ is the electron mass. The proximity-induced gap can be within the range $\Delta=0.2$-$1~{\rm meV}$ as discussed in Table II of Ref.~\cite{lut18} and here we choose $\Delta=0.5~{\rm meV}$. The length scale of the proximitized superconducting gap generally depends on the diffusive constant of the wire and temperature. However, microscopic calculations for InAs nanowires with an epitaxial aluminum (Al) shell demonstrate that the proximitized gap is comparable to that of the parent superconductor and extends over the entire wire with a realistic thickness~\cite{winkler19}. Therefore, we here assume $\Delta$ to be constant.

We are interested in the signatures of the TPT in the density response of Eq.\,(\ref{eq:H}). For this reason, we first inspect the TPT, which can be identified from the quasiparticle excitation energies of Eq.\,(\ref{eq:H}) obtained as
\begin{align}
E^2_{\pm}(k)
=P_k 
 \pm 2 \sqrt{b^2\Delta^2+\xi^2_k(b^2+\alpha^2k^2)}.
 \label{eq:Ek}
\end{align}
where $P_k=\Delta^2+\xi^2_k+b^2+\alpha^2k^2$. We first note that, in the normal state ($\Delta=0$), there are two Fermi points, $k^{\rm N}_{{\rm F},\pm}$. The field evolution of the energy dispersions [Eq.\,(\ref{eq:Ek})] in the normal state is displayed in Fig.~\ref{fig:setup0}(b). The inner Fermi point ($k_{{\rm F},+}^{\rm N}$) approaches zero with increasing $B$, reaching zero at $B_{\rm L}\equiv\mu= 0.69~{\rm T}$ and marking the Lifshitz transition also known as helical transition \cite{cay15}. To understand the superconducting state, Fig.~\ref{fig:setup0}(c) displays  $E_{\pm}(k)$ as a function of $k$ at distinct $B$. Here, $B$ causes the bulk excitation gap to close, $E_{-}(k=0)\rightarrow 0$, at the critical field $B_{\rm c} \equiv \sqrt{\mu^2+\Delta^2}$. This signals the TPT from the trivial ($B<B_{\rm c}$) to the topological phase ($B>B_{\rm c}$), see Fig.~\ref{fig:setup0}(c). For the chosen parameters, $B_{\rm c} =0.70~{\rm T}$, which deviates from $B_{\rm L}$.

\begin{figure}[t!]
\includegraphics[width=85mm]{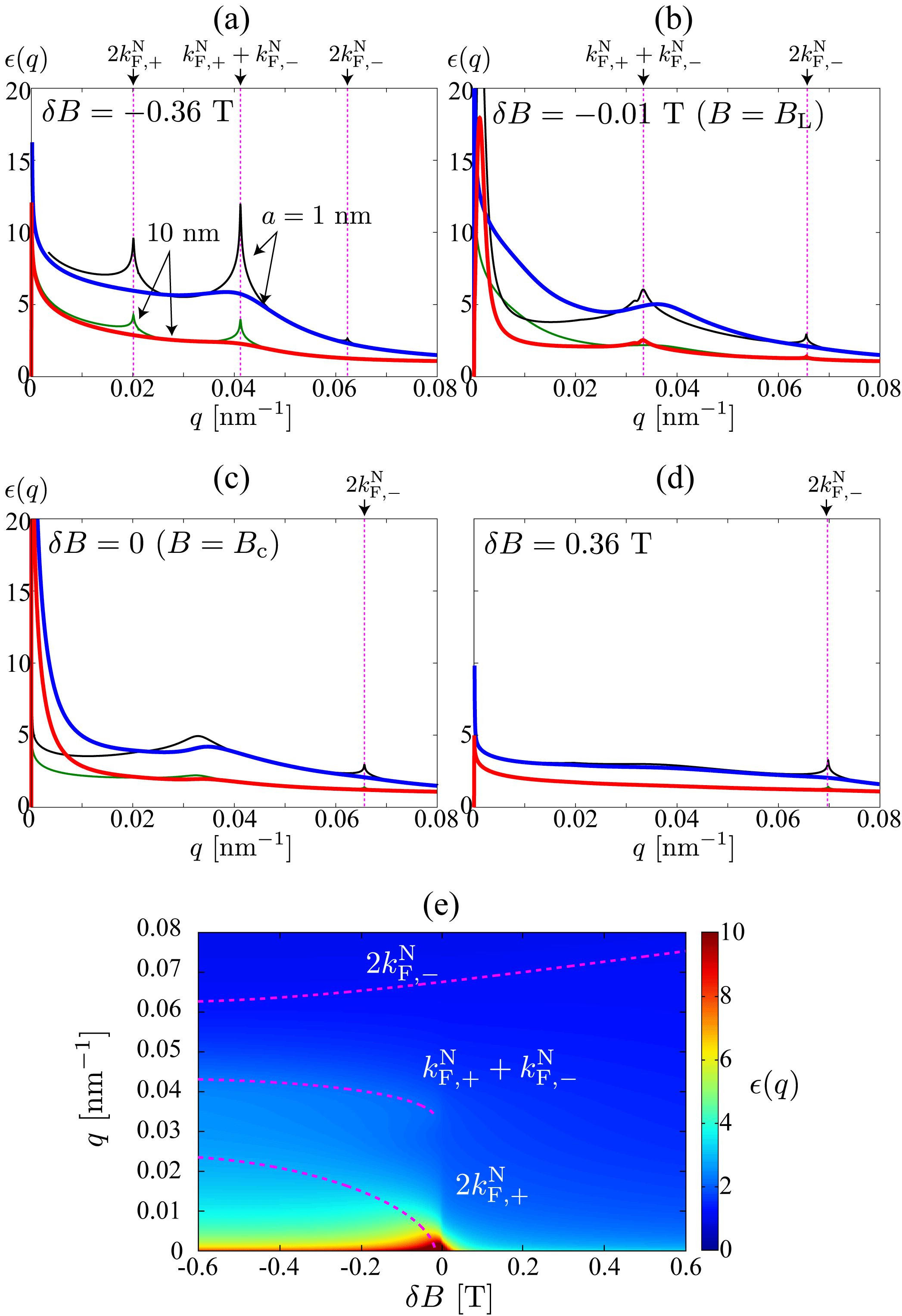}
\caption{Static dielectric function, $\epsilon(q)$, in the superconducting state (thick red/blue curves) and in the normal state (thin black/green curves) at $\delta B \equiv B-B_{\rm c}=-0.36~{\rm T}$ (a), $\delta B=-0.01~{\rm T}$ ($B=B_{\rm L}$) (b), $\delta B=0$ ($B=B_{\rm c}$) (c), and $\delta B=0.36~{\rm T}$ (d), where we set $T=0.01T_{\rm c}$. The thick blue and red (thin black and green) curves correspond to $\epsilon(q)$ in the superconducting (normal) state with the radius $a=1~{\rm nm}$ and $10~{\rm nm}$, respectively. (e) Field dependence of $\epsilon(q)$ in the superconducting state. The dotted lines show the $B$-dependences of the characteristic wavenumbers, $2k^{\rm N}_{{\rm F},\pm}$ and $k^{\rm N}_{{\rm F},+}+k^{\rm N}_{{\rm F},-}$.}
\label{fig:dd}
\end{figure}

\section{Static charge screening}

We start by examining the signatures of the TPT in the static density response and charge screening. The bare dynamic density-density correlation function is defined with the density operator, $\hat{\rho}({\bm q}) \equiv \sum_{{\bm k},\sigma}c^{\dag}_{{\bm k}+{\bm q},\sigma}c_{{\bm k},\sigma}$ and the volume of the system, $V$, as \cite{giu}
\beq
\chi_{\rho\rho}(Q) = -\frac{1}{V}\int^{T^{-1}}_0 
\left\langle
\hat{\rho}({\bm q},\tau)\hat{\rho}(-{\bm q},0)
\right\rangle e^{i\omega_m\tau}d\tau.
\label{eq:chinn}
\eeq
We set $Q\equiv({\bm q},i\omega_m)$, where $\omega_m=2m\pi T$ is the Matsubara frequency for bosons at temperature $T$ ($m\in \mathbb{Z}$). The function is then expressed in terms of the equilibrium Green's functions, $G(K)\equiv G(k,i\varepsilon_n)= [i\varepsilon_n-\mathcal{H}(k)]^{-1}$, as 
$\chi_{\rho\rho}(Q) = \sum_K{\rm tr}_4[
G( K)
\tau_z
G( K+Q)
\tau_z]$,
where $\varepsilon_n = (2n+1)\pi T$ is the Matsubara frequency for fermions ($n\in \mathbb{Z}$), $\sum_K \equiv T\sum_n\frac{1}{V}\sum_{{\bm k}}$, and ${\rm tr}_4$ denotes the trace over Nambu and spin spaces. For the expression of $G(k)$, see Sec.~\ref{sec:G}. We consider the long-range Coulomb interaction between electrons in addition to the effective Hamiltonian of Eq.~\eqref{eq:H}, representing the backflow to ensure particle number conservation of charge density fluctuations. The Coulomb potential in a cylindrical wire of radius $a$ is given by $U(q) = -\frac{e^2}{4\pi}e^{q^2a^2} {\rm Ei}( -q^2a^2 )$ with the exponential-integral function ${\rm Ei}(x) = -\int ^{\infty}_{-x} \frac{e^u}{u}du$~\cite{cayao_phd,giu}. Within the random phase approximation, the generalized dynamic density-density response function is obtained as \cite{giu}
\begin{align}
\tilde{\chi}_{\rho\rho}(q,\omega) = \frac{\chi_{\rho\rho}(q,\omega)}{1-U(q)\chi_{\rho\rho}(q,\omega)}.
\label{eq:chi_rpa}
\end{align}
where $i\omega_{m}\rightarrow\omega+i\eta$. To capture the signal of the bulk gap closing and reopening from the static density response, we calculate the dielectric function, $\epsilon(q) = 1-U(q){\chi}_{\rho\rho}(q,\omega=0)$ and show it in Fig.~\ref{fig:dd} in the normal and superconducting states.

 In the normal state, the static dielectric function $\epsilon(q)$ as a function of $q$ is depicted by thin black and green curves in Figs.~\ref{fig:dd}(a-d), which correspond to a radius of $a=1$\,nm and $a=10$\,nm, respectively. The most inmediate feature to notice is that $\epsilon(q)$ develops three resonance peaks at $q=\{2k_{{\rm F},+}^{\rm N}, k_{{\rm F},+}^{\rm N} + k_{{\rm F},-}^{\rm N}, 2k_{{\rm F},-}^{\rm N}\}$. These peaks represent the typical dispersions of normal electrons, which contrast with the case of free electrons in which the density response function exhibits only a single peak at $q=2k_{\rm F}$ \cite{giu}. The visibility of the peaks diminishes with increasing $a$ and is washed out when $q>2\pi/a$. At the Lifshitz transition ($B=B_{\rm L}$), the chemical potential touches the bottom of the inner band and the number of the Fermi surfaces reduces from two to one [Fig.\,\ref{fig:setup0}(b)]. In this case at $B=B_{\rm L}$, the dielectric function diverges at $q\rightarrow 0$, see thin black and green curves in Fig.~\ref{fig:dd}(b). As $B$ further increases, the peak at $q\rightarrow 0$ dimishes, as observed in Figs.~\ref{fig:dd}(c) and \ref{fig:dd}(d).

In the superconducting state ($\Delta\neq 0$), the field evolution of $\epsilon(q)$ captures noteworthy features of the bulk excitations. In the trivial phase, the three singular peaks weaken since the superconducting order gaps out the Fermi points of normal electrons, see Figs.~\ref{fig:dd}(a). Even though these peaks are washed out, a singular behavior develops at $q=0$ when the TPT takes place at $B=B_{\rm c}$ [Fig.~\ref{fig:dd}(c)]. This is attributed to the fact that the lowest energy of the pair excitations in the inner band, $\omega^+_{\rm pair}\equiv\min[E_-(k_{{\rm F},+}^{\rm N})+E_-(-k_{{\rm F},+}^{\rm N})]$, becomes zero at the TPT, significantly contributing to the bare density response function in the static limit, $\chi_{\rho\rho}(q,\omega=0)$. We note that the singular behavior of $\epsilon(q)$ is insensitive to the radius of the wire, $a$, while it disappears in the topological phase [Fig.~\ref{fig:dd}(d)]. Figure~\ref{fig:dd}(e) summarizes the field evolution of the dielectric function. The dielectric function in the long wavelength limit is considerably enhanced as $B=B_{\rm c}$ is approached, signaling the bulk gap closing.

Before going further, we emphasize that the static dielectric function discussed above,  $\epsilon(q)$, can be  accessed by means of compressibility. This is because $\epsilon(q)$   can be directly related to the compressibility, $\kappa \equiv \frac{1}{n^2}\frac{\partial n}{\partial \mu}$, through the sum rule of the static density response function as~\cite{giu}
\begin{align}
\lim _{q\rightarrow 0} \epsilon(q) =  1 + U(q) \frac{\partial n}{\partial \mu},
\end{align}
where $n$ is the density of electrons. In this regard, we note that capacitive techniques have been used to measure the compressibility of low-dimensional semiconductors~\cite{eisen92,eisen94,cast98,lus07,smith11}. Therefore, the evolution of the characteristic peak structures across the TPT may be observed in compressibility measurements~\cite{nozadze16}.

\section{Dynamical response and Raman spectra}

\begin{figure}[t!]
\includegraphics[width=85mm]{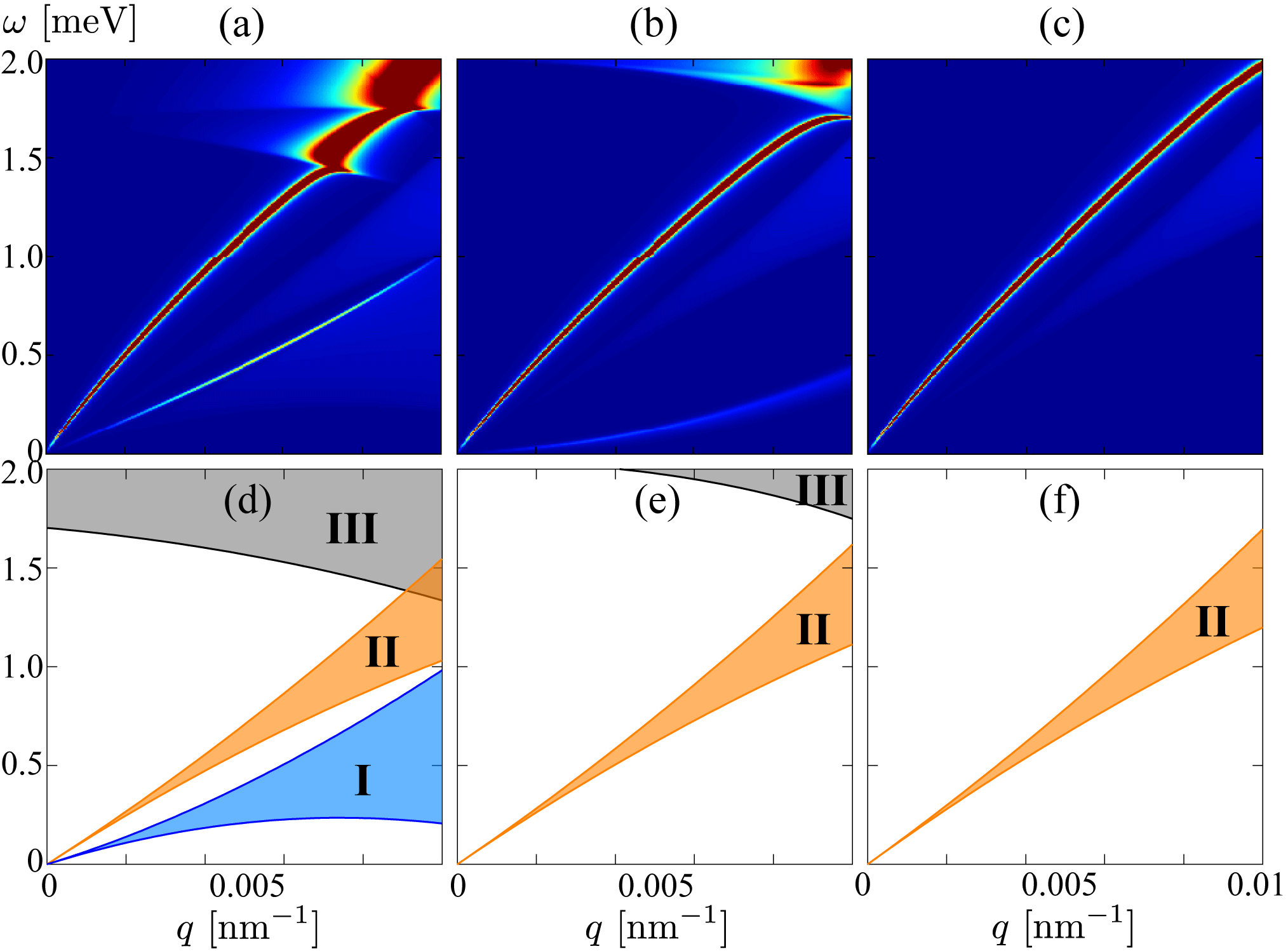}
\caption{(a-c) Density response $-{\rm Im}\tilde{\chi}_{\rho\rho}(q,\omega)$ in the normal state $\Delta=0$ at $\delta B = -0.18~{\rm T}$ (a), $0$ (b), and $0.18~{\rm T}$ (c), where $\delta B = 0$ corresponds to the Lifshitz transition since here $\Delta=0$. (d-f) Three particle-hole continua are denoted by processes I, II, and III, where I (II) is for intraband excitations in the inner (outer) band, while process III corresponds to the interband excitations. We take $a=10~{\rm nm}$ and the other parameters are the same as those in Fig.\,\ref{fig:dd}.}
\label{fig:normald}
\end{figure}

Having explored signatures of the TPT in the static dielectric function, here we extend the analysis into the dynamical density response which is crucial in spectroscopic measurements~\cite{altland2010condensed,giu}.

\subsection{Dynamical response in the normal state of the semiconducting wire}
We first consider the dynamical density response in the normal semiconducting wire described by Eq.~\eqref{eq:H} with $\Delta=0$. In the normal wire, applying the magnetic field causes the Lifshitz transition at $B_{\rm L}$ above which the inner Fermi surface disappears. Since this critical field always appears just below the critical field in the superconducting state, $B_{\rm c}=\sqrt{B^2_{\rm L}+\Delta^2}$, it serves as a precursor to the topological phase transition \cite{cay15}.

Figure~\ref{fig:normald} shows the evolution of the renormalized density response function from Eq.\,(\ref{eq:chi_rpa}), $-{\rm Im}\tilde{\chi}_{\rho\rho}(q,\omega)$, at $\Delta=0$ and for a narrow cylinder with Coulomb potential $U(q)\approx -\frac{e^2}{2\pi}\ln{(qa)}$ in $qa\ll 1$. As shown in Fig.~\ref{fig:normald}(a), there exist two gapless plasmon modes at $\delta B=-0.18~{\rm T}$, which then soften as $B$ passes the Lifshitz transition at $B_{\rm L}$ [Fig.~\ref{fig:normald}(b)] and only one plasmon mode remains in the helical phase $B>B_{\rm L}$ [Fig.~\ref{fig:normald}(c)].  The lower (upper) branch corresponds to the plasma mode originating from the inner (outer) band with the Fermi points at $\pm k^{\rm N}_{{\rm F},+}$ ($\pm k^{\rm N}_{{\rm F},-}$). Further insights are obtained from Figs.~\ref{fig:normald}(d-f), where, along with the sharp peaks of Figs.~\ref{fig:normald}(a-c) we find three particle-hole continua are denoted by regions I, II, and III. Regions I and II correspond to the particle-hole continua in the inner and outer bands, respectively, while region III represents where interband excitations occur (see Appendix~\ref{sec:wiren}). As seen in Figs.~\ref{fig:normald}(d-f), the upper plasmon branch, which exists above region II, is the one remaining in the helical phase after the Lifshitz transition. Gapless plasmons generally appear in one dimensional electrons~\cite{sar85,li90,sat93,sar96}, which were detected by inelastic Raman scattering in semiconducting wires~\cite{ege90,gon91,boo03}. Thus, the field evolution of gapless plasmons offers a robust way to identify the bulk Lifshitz transition in semiconducting wires.

\begin{figure}[t!]
\includegraphics[width=85mm]{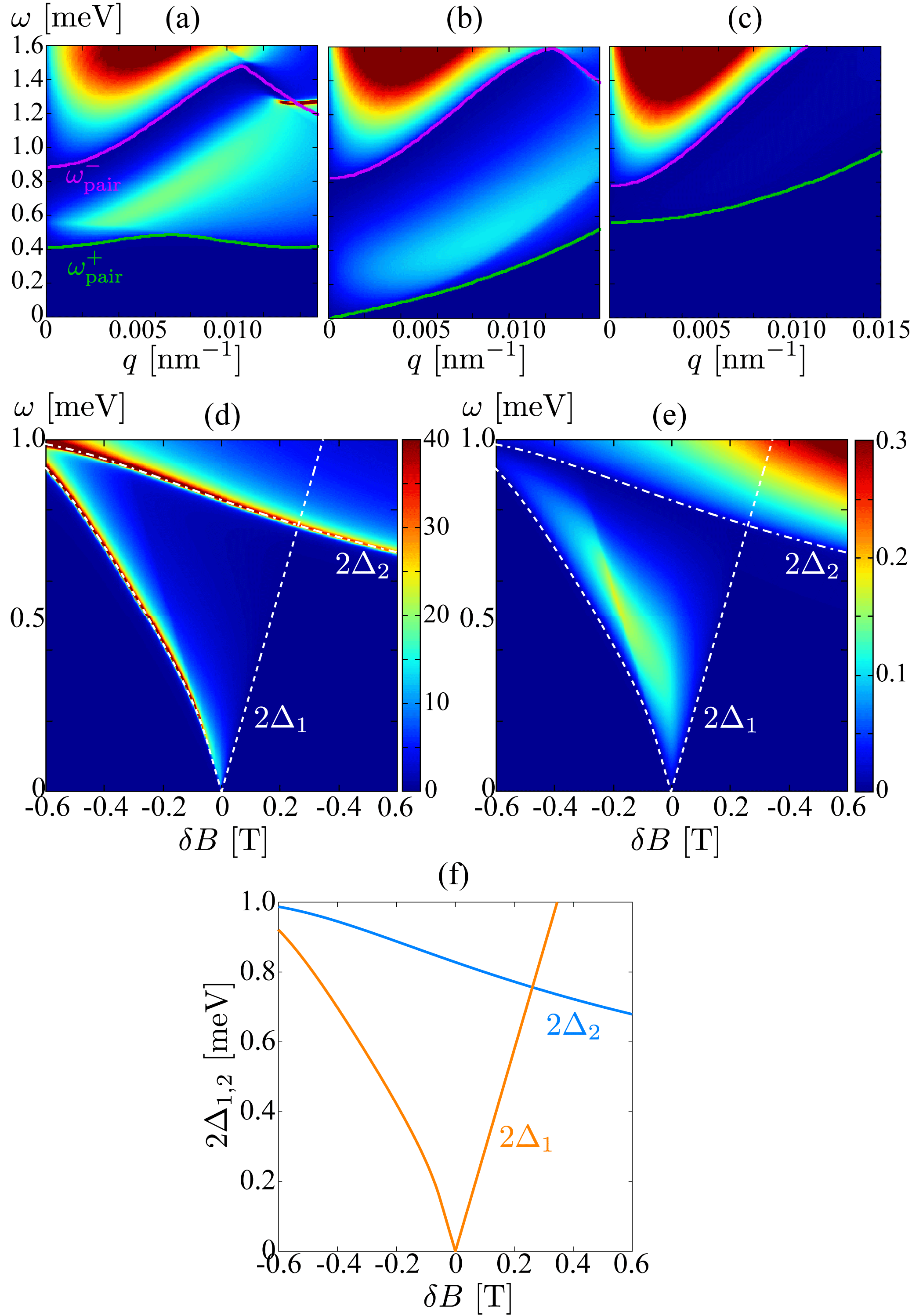}
\caption{(a-c) Renormalized density response $-{\rm Im}\tilde{\chi}_{\rho\rho}(q,\omega)$ in the superconducting state as a function of $q$: (a) $\delta B = -0.18~{\rm T}$, (b) $0$ ($B=B_{\rm c}$), and (c) $0.18~{\rm T}$. The green (magenta) curve denotes $\omega^+_{\rm pair}$ ($\omega^-_{\rm pair}$), which is the lowest pair excitation energies within the inner and outer bands, respectively [see Fig.~\ref{fig:setup0}(b)]. (d) Bare density response, $-{\rm Im}\chi_{\rho\rho}(q,\omega)$ and (e) renormalized density response $-{\rm Im}\tilde{\chi}_{\rho\rho}(q,\omega)$ as a function of $B$ at $q=1.0\times 10^{-4}~{\rm nm}^{-1}$. In (d,e), the thin dashed and dotted-dashed lines correspond to the field dependence of the superconducting gaps, $2\Delta_1$ and $2\Delta_2$, respectively. We remark that $\tilde{\chi}_{\rho\rho}$ coincides with the Raman response function, $\tilde{\chi}_{\gamma\gamma}$, at $\gamma = 1$. Here we set $a=10~{\rm nm}$. (f) Quasiparticle excitation gaps in the inner and outer Fermi surfaces, $2\Delta_{1,2}$, as a function of $B$. See Fig.~\ref{fig:setup0}(c) for the definition of $\Delta_{1,2}$.}

\label{fig:disp}
\end{figure}

\subsection{Superconductor-semiconductor hybrids}

In the superconducting state, Figs.~\ref{fig:disp}(a-c) show the renormalized density response $-{\rm Im}\tilde{\chi}_{\rho\rho}(q,\omega)$ obtained from Eq.\,(\ref{eq:chi_rpa}) as a function of $q$. We would like to emphasize that $\Delta$ in Eq.~\eqref{eq:H} is the proximity-induced gap from the parent superconductor and not the self-consistent field within the semiconducting wire. If the wire is intrinsically superconducting on its own, it is essential to consider the phase fluctuation of the superconducting order to ensure a gauge-invariant response. Under these circumstances, the gapless plasmon modes persist even in the superconducting state and the density response changes from Fig.~\ref{fig:disp}. However, the hybrid system we consider here is not the case. Indeed, as shown in Figs.~\ref{fig:disp}(a-c), the density response in the superconducting state does not have the prominent peaks associated with plasmons that exist in the normal state in Fig.~\ref{fig:normald}, leaving behind two continua in the low frequency region whose field evolution traces the bulk gap closing and reopening across the TPT. As shown in Fig.~\ref{fig:disp}(a), two excitation continua appear in the trivial phase $B<B_{\rm c}$, with threshold energies $\omega^{\pm}_{\rm pair}(q)$ depicted by green and magenta curves. Here, $\omega^{\pm}_{\rm pair}(q)$ are defined as $\omega^{\pm}_{\rm pair}(q)={\rm min}[E_{\pm}(k+q)+E_{\pm}(-k)]$ and represent the lowest energy of intraband pair excitations. Because of the large difference between two Fermi momenta, e.g., $k^{\rm F}_-- k^{\rm F}_+\approx 0.025~{\rm nm}^{-1}$ at $\delta B=-0.18~{\rm T}$, the interband pair-breaking continuum only appears at large $q$ values outside of Fig.~\ref{fig:disp}. At the TPT, $B=B_{\rm c}$, in Fig.~\ref{fig:disp}(b), the threshold energy of the lower continuum with $\omega_{\rm pair}^{+}$ goes to zero, while the upper continuum with $\omega^-_{\rm pair}$ remains less sensitive across the TPT. Once the bulk gap is reopened in the topological phase, the lower continuum vanishes, and only the upper continuum contributes to the density response [Fig.~\ref{fig:disp}(c)]. Thus, the field evolution of the dynamical density response traces the bulk gap closing and reopening across the TPT.

As shown in Fig.~\ref{fig:disp}(c), in the topological phase, the density response in the frequency range from $\omega^{+}_{\rm pair}$ to $\omega^-_{\rm pair}$ is vanishingly weak. Similarly, in Figs.~\ref{fig:disp}(d) and \ref{fig:disp}(e), the intensity of the lower continuum for $\delta B >0$ (the region within $2\Delta_1<\omega<2\Delta_2$) is weaker than that for $\delta B<0$. These imply that pair excitations from the inner band become considerably weaker across the TPT. We attribute this suppression of the pair excitations to a loss of superconducting coherence in the inner band. When focusing on the inner band, the band bottom rises above the Fermi energy at the TPT, and the inner Fermi surface vanishes in the topological phase. The disappearance of the inner Fermi points inhibits the formation of Cooper pairs, leading to a loss of superconducting coherence, in particular, when the bottom of the inner band is more than $\Delta=0.5~{\rm meV}$ away from the Fermi energy. This suppression of superconducting coherence also reduces pair excitations from the inner band in $\delta B>0$.

The effective dynamical density fluctuations discussed in the previous paragraph can be measured by electronic Raman spectroscopy. In fact, the Raman scattering intensity of incident light is represented by the differential photon cross-section that measures the probability of the scattering into a solid angle ($d\Omega$) and an energy window ($d\omega$). 
The differential cross section is given by~\cite{klein84,dev07}
\begin{align}
\frac{\partial^2\sigma}{\partial \Omega\partial\omega}
=\frac{\omega_{\rm S}}{\omega_{\rm I}}r^2_0S({\bm q},\omega),
\label{eq:R}
\end{align}
where $\omega = \omega _{\rm I}-\omega_{\rm S}$ and ${\bm q}={\bm q}_{\rm I}-{\bm q}_{\rm S}$ are the energy and wave vector differences between the incident $({\bm q}_{\rm I},\omega_{\rm I})$ and scattered light $({\bm q}_{\rm S},\omega_{\rm S})$, respectively, and $r_0=e^2/mc^2$ is the Thompson radius. The distribution function of bosons at temperature $T$ is given by $f_{\rm B}(\omega)=1/(e^{\omega/T}-1)$. The generalized structure function $S$ in Eq.~\eqref{eq:R} is related to the imaginary part to the Raman response function $\tilde{\chi}_{\gamma\gamma}$~\cite{klein84,mon90,dev07},
$S({\bm q},\omega) = -\frac{1}{\pi}(
1+f_{\rm B}(\omega)
){\rm Im}\tilde{\chi}_{\gamma\gamma}({\bm q},\omega)$,
where the Raman response function can be written as
\begin{align}
\tilde{\chi}_{\gamma\gamma} (Q)= \chi_{\gamma\gamma} (Q)
+ \chi_{\gamma \rho}(Q) \mathcal{D}^{\rm w}_{\rho}(Q)\chi_{\rho\gamma}(Q).
\label{eq:chi_gamma}
\end{align}
The functions $\chi_{\gamma \rho}(Q)=\chi_{\rho\gamma}(-Q)$ and $\chi_{\gamma\gamma}$ describe the density-Raman correlation and Raman density-Raman density correlation, respectively, which are obtained from Eq.~\eqref{eq:chinn} by replacing $\hat{\rho}({\bm q})$ with the Raman density operator $\hat{\rho}_{\rm R}({\bm q}) \equiv \sum_{{\bm k},\sigma}\gamma_{\bm k}c^{\dag}_{{\bm k}+{\bm q},\sigma}c_{{\bm k},\sigma}$. The second term in Eq.~\eqref{eq:chi_gamma} contains the propagator of the charge fluctuation, $\mathcal{D}^{\rm w}_{\rho}(Q)\equiv [U^{-1}(q) - \chi_{\rho\rho}(Q)]^{-1}$, representing the backflow to ensure particle number conservation. Since we consider a parabolic band, the Raman vertex is $\gamma = 1$ and the Raman response given by Eq.~\eqref{eq:chi_gamma} coincides with the renormalized density response given by $\tilde{\chi}_{\rho\rho}$ in Eq.~\eqref{eq:chi_rpa}. This behavior remains as long as $\gamma$ is constant;  the $k$-dependence of $\gamma$ arises due to nonparabolic bands, but it induces a small correction to  Eq.~\eqref{eq:chi_gamma}. Thus, we next focus on $\tilde{\chi}_{\rho\rho}$ instead of $\tilde{\chi}_{\gamma\gamma}$.

The field evolution of the bare and renormalized density response functions, $-{\rm Im}\chi_{\rho\rho}(q,\omega)$ and $-{\rm Im}\tilde{\chi}_{\rho\rho}(q,\omega)$, at $q=1\times 10^{-4}~{\rm nm}^{-1}$ is displayed in Figs.~\ref{fig:disp}(d) and \ref{fig:disp}(e). In Raman scattering experiments, $q=q_{\rm I}-q_{\rm S}$ is the momentum transferred from the incident light, and its scale reflects the wavelength of photons. In the THz frequency band, the wavelength is of the order of $\lambda_{\rm ph}\equiv 2\pi/q_{\rm ph}\sim 100~\mu{\rm m}$. The magnitude of the transferred momentum, $q$, is controlled by the incidence angle $\theta$, i.e., $q=q_{\rm ph}\cos\theta = O( 10^{-4}~{\rm nm}^{-1})$. Figure~\ref{fig:disp}(d) reveals that two sharp peaks develop as a function of $B$ in the bare density response. In Figs.~\ref{fig:disp}(d-f), we also plot the field dependence of $2\Delta_{1}$ and $2\Delta_2$, corresponding to the lowest energy of the pair excitations in the inner and outer bands, respectively. Notably, the lowest peak traces the field evolution of the lowest energy of the pair excitations in the inner band, $\omega^+_{\rm pair}(0)=E_+(k^{\rm N}_{{\rm F},+})+E_+(-k^{\rm N}_{{\rm F},+})=2\Delta_1$, capturing the bulk gap closing and reopening at the TPT $B=B_{\rm c}$. The upper branch remains less sensitive to $B$, implying that the superconducting gap at the outer Fermi point ($k=k^{\rm N}_{{\rm F},-}$) is finite across the TPT. The charge fluctuation at the long wavelength is screened by the long-range Coulomb interaction, and the renormalized density response, $\tilde{\chi}_{\rho\rho}$, reduces with $q^2$ for $q\rightarrow 0$.  As shown in Fig.~\ref{fig:disp}(e), the intensity of $\tilde{\chi}_{\rho\rho}$ is suppressed by the screening effect but still captures the bulk gap closing and reopening at $B_{\rm c}$. This result is not affected by the value of $\Delta$, see Appendix.~\ref{sec:wires}. We can thus conclude that the field evolution of the Raman response function detects the TPT in superconductor-semiconductor hybrids.

\begin{figure}[t!]
\includegraphics[width=85mm]{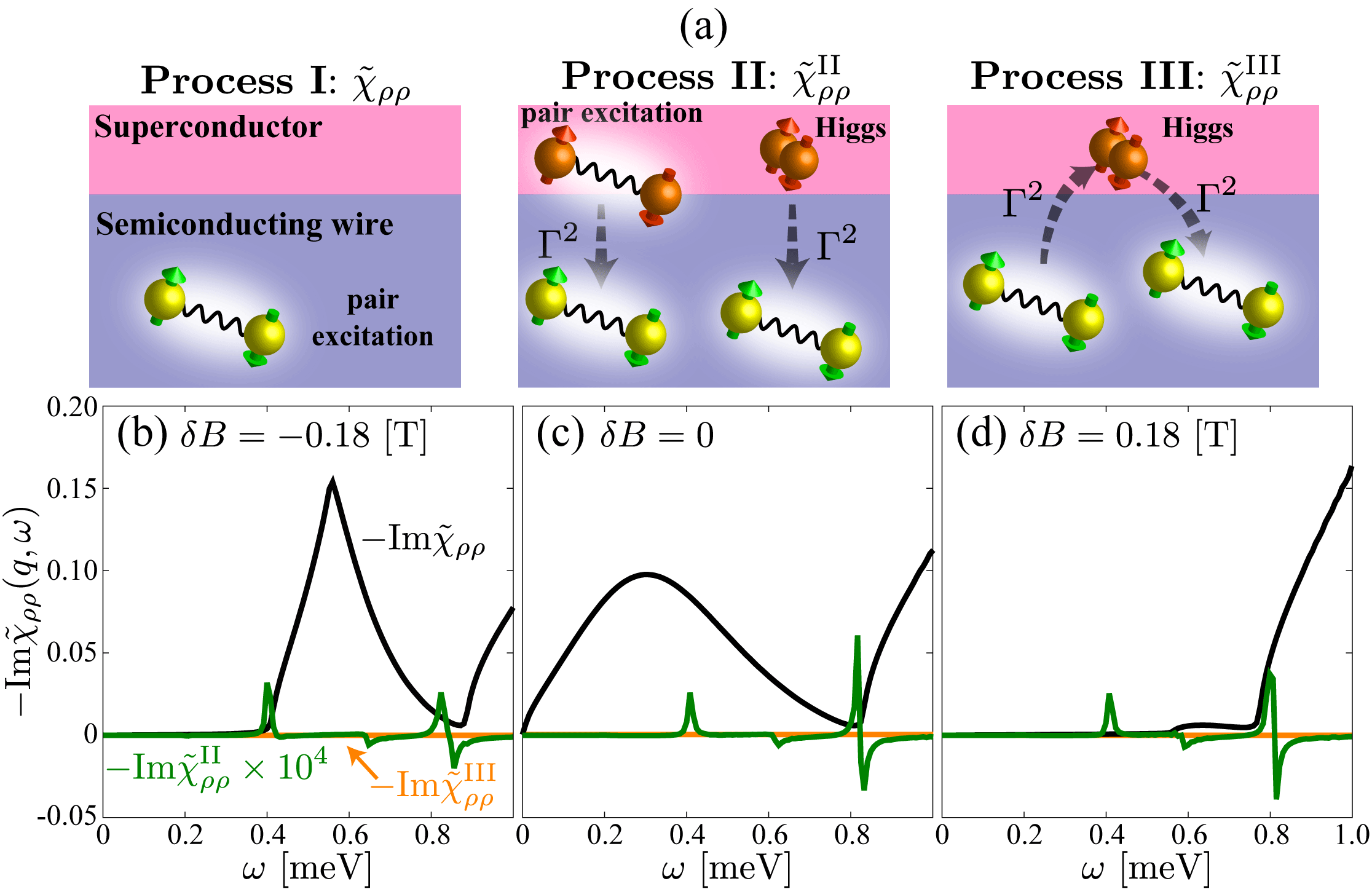}
\caption{(a) Three processes that influence the density response of the semiconducting wire. Process I corresponds to $\tilde{\chi}_{\rho\rho}$ in Eq.~\eqref{eq:chi_rpa}.  Process II involves the tunneling of pair excitations and Higgs bosons driven by an incident light into the wire, while process III is mediated by collective excitations in the parent superconductor. (b,c) Density response functions through three processes: (b) $\delta B=-0.18~{\rm T}$, (c) $0$, and (d) $0.18~{\rm T}$ at $q=1.0\times 10^{-4}~{\rm nm}^{-1}$. Here we take the tunneling energy as $\Gamma={\mu}=0.2~{\rm meV}$. We also choose the gap and the chemical potential of the parent superconductor as $\Delta_{\rm sc}=1~{\rm eV}$ and $\mu_{\rm sc}=0.2~{\rm meV}$, respectively (see Sec.~\ref{sec:CM}).}
\label{fig:cm}
\end{figure}

\subsection{Collective modes of the parent superconductor}

To further support the detection of the TPT via Raman spectroscopy, we now address the fluctuation effect in the parent superconductor. Let us consider the microscopic action, which is composed of a semiconducting wire coupled to a conventional superconductor $\Delta_{\rm sc}$, where electrons can tunnel between the two subsystems with an energy $\Gamma$. Here we assume the spatial uniformity of the parent superconductor. In practical experiments, the superconducting film must be sufficiently thin to endure high in-plane magnetic fields without losing its superconductivity. In fact, in epitaxially grown Al on semiconducting nanowires, the gap size increases as the film becomes thinner, and a sharp U-shaped density of states was observed through conductance measurements~\cite{chang15}. These ideas allow one to assume the parent superconductor as a spatially uniform bulk system.

The system comprises two subsystems: the semiconducting wire and the parent superconductor. In this case, the dynamical response of the semiconducting wire to an external perturbation involves the following three processes, as shown in Fig.~\ref{fig:cm}(a).
%
%
Process I is the direct response of the semiconducting wire, corresponding to  $\tilde{\chi}_{\rho\rho}$ in Eq.~\eqref{eq:chi_rpa}. In process II, both pair excitations and collective excitations, including the Higgs mode, are driven in the parent superconductor by an incident light, which then tunnel into the wire. Process III is mediated by the collective excitations in the parent superconductor, which renormalizes the charge fluctuation propagator in the wire. Therefore, the dynamical density response of the wire in total is represented by the sum of these three processes, $\tilde{\chi}_{\rho\rho}^{\rm tot}= \tilde{\chi}_{\rho\rho}+\tilde{\chi}_{\rho\rho}^{\rm II}+\tilde{\chi}_{\rho\rho}^{\rm III}$, see Appendix~\ref{sec:CM} for the detailed derivations and expressions for $\tilde{\chi}_{\rho\rho}^{\rm II,III}$.

Figs.~\ref{fig:cm}(b-d) show the density response functions of the wire through these three processes at $\delta B=-0.18~{\rm T}$, 0, and $0.18~{\rm T}$, respectively, where we set $\Gamma = \mu$, where $\mu$ is the chemical potential of electrons in the wire. All the parameters in the wire are the same as those used before, while the gap and chemical potential in the parent superconductor are chosen as $\Delta_{\rm sc}=0.2~{\rm meV}$ and $\mu_{\rm sc}=1~{\rm eV}$, respectively. It is observed that the density response of the wire is dominated by process I corresponding to Eq.~\eqref{eq:chi_rpa}, see black curve in Fig.~\ref{fig:cm}(b). In the process II ($\chi^{\rm II}_{\rho\rho}$), a tiny peak appears at $\omega = 0.4~{\rm meV}$, which coincides with the threshold energy of pair excitations and Higgs excitation energy in the parent superconductor, see green curve in Fig.~\ref{fig:cm}(b). 
We have verified that, although the peak slightly increases with $\Gamma$, the contributions through processes II and III remain negligible compared to process I also at the TPT and the topological phase [Figs.~\ref{fig:cm}(c) and \ref{fig:cm}(d)]. We remark that, in the weak coupling limit when the gap of the parent superconductor is much smaller than its chemical potential, the density response function of the parent superconductor vanishes due to the screening effect of the long-range Coulomb interaction~\cite{cea14,cea16}, see also Appendix~\ref{sec:sc_cm} and Fig.~\ref{fig:higgs}(c). Although the coupling to the wire via pair tunneling slightly diminishes the screening effect, the density response of the parent superconductor remains negligible, at least when $\Gamma \lesssim \mu_{\rm sc}= 1~{\rm meV}$.
In contrast, as shown in Fig.~\ref{fig:chiq0}, the intensity of the density response in a superconducting nanowire increases with its radius $a$, indicating that increasing the wire’s radius weakens the screening effect. This is because electrons in the wire with a larger $a$ have more space to move and hence their pair-wise interaction is weaker.
Therefore, the layer of the parent superconductor provides little contribution to the density response and the structure factor in Eq.~\eqref{eq:R} is dominated by the dynamic density response of the semiconducting wire in Eq.~\eqref{eq:chi_rpa}. The Raman response functions connected with the superconducting layers, $\tilde{\chi}^{\rm II,III}_{\rho\rho}$, do not contribute significantly. The field evolution of the Raman response shown in Fig.~\ref{fig:disp} can detect the TPT even when considering the fluctuation effects in the parent superconductor.

\section{Conclusion}

We have studied how to identify a signature of topological bulk gap closing and reopening through the density response of superconductor-semiconductor hybrids. In particular, we have demonstrated that the dynamical density-density response functions reveal the field evolution of the bulk excitation gap across the topological phase transition. We have shown that the dynamical response signatures of the topological phase transition can be measured by electronic Raman spectroscopy. Furthermore, we have found that the normal state of the semiconducting wire hosts two gapless plasmons, and their field evolution provides a way to identify the Lifshitz transition. Inelastic Raman scattering in similar systems was already used to study plasmonic excitations~\cite{ege90,gon91,boo03}. Our findings, therefore, offer an alternative way for measuring the topological phase transition in superconductor-semiconductor hybrids.

We note that, even when trivial ABSs form at the interface, bulk measurements like compressibility measurements and Raman spectroscopy may capture signals of the topological phase transition. As discussed in Ref.~\cite{hes23} (see also the discussions in Refs.~\cite{sar24,hes24}, however, the trivial ABSs may come into play in such bulk measurements if an Andreev band forms along the entire wire. The wave functions of this Andreev band may extend across the whole wire if there is a periodic or quasi-periodic distribution of disorders, which can obscure the signal of the bulk topological phase transition. But, the issue of how can bulk measurements distinguish between genuine bulk signals and those from trivial Andreev bands requires a separate proper investigation and we expect to address it in a future work.

\begin{acknowledgments}  
We thank J. A. Castellanos-Reyes for useful discussions. 
T. M. and Y. T. acknowledge financial support from
JSPS with Grants-in-Aid for Scientific Research (KAKENHI Grants No.~JP21H01039, No.~JP22H01221, No.~JP23K17668, No.~JP23K20828, No.~JP23K22492, No.~JP24K00583, No.~JP24K00556, No.~JP24K00578, No.~JP25H00599, No.~JP25H00609, No.~25K07227, and No.~JP25K22011) and a Grant-in-Aid for Transformative Research Areas (A) ``Correlation Design Science'' (Grant No.~JP25H01250). 
J. C. acknowledges financial support from the Swedish Research Council (Vetenskapsr{\aa}det Grant No. 2021-04121), the Sweden-Japan Foundation (Grant No. BA24-0003), and the Carl Trygger’s Foundation (Grant No. 22: 2093).
\end{acknowledgments}


\appendix

\section{Green's function of the Rashba superconductor}
\label{sec:G}

\begin{figure*}[t!]
\includegraphics[width=160mm]{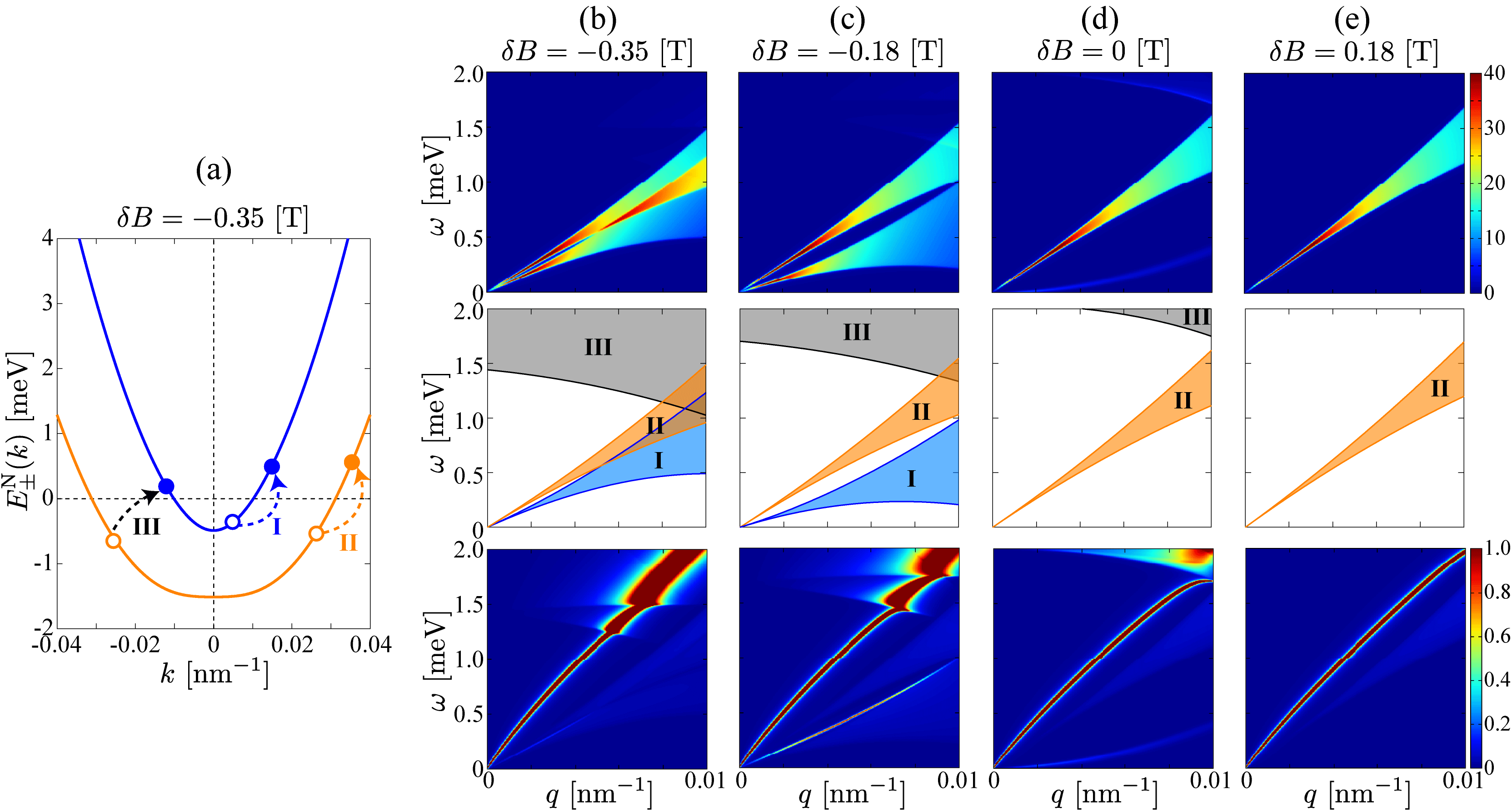}
\caption{Excitation spectra of normal electrons in the semiconducting wire ($\Delta_0^{\rm s}=0$). (a) Dispersion of the normal electrons in the wire at $\delta B = -0.35~{\rm T}$ and three particle-hole pair excitations (I, II, and III), where the process I (II) is the excitations in the inner (outer) band and the process III is the interband excitation. (b-e) (Lower panels) Spectra of the bare density-density response function in the semiconducting wire, ${\rm Im}\chi^{\rm ww}_{11}(q,\omega)=-{\rm Im}\chi_{\rho\rho}(q,\omega)$: (a) $\delta B = -0.35~{\rm T}$, (b) $-0.18~{\rm T}$, (c) $0$, and $0.18~{\rm T}$, where $\delta B$ corresponds to the Lifshitz transition. (Middle panels) Particle-hole continua corresponding to the process I, II, and III are depicted by the shaded area. (Lower panels) Spectra of the Coulomb propagator in the semiconducting wire, $-{\rm Im}\mathcal{D}^{\rm ww}_{11}(q,\omega)$. The gapless branches are the gapless plasma modes. At $\delta B=-0.18~{\rm T}$, another gapless plasma mode appears in the frequency region between the continua I and II, which becomes softening at $\delta B=0$ and disappears in $\delta B>0$. }
\label{fig:normal}  
\end{figure*}

We consider a one-dimensional superconductor with Rashba spin-orbit coupling. The Bogoliubov-de Gennes (BdG) Hamiltonian in the basis $(c_{k,\uparrow},c_{k,\downarrow},c^{\dag}_{-k,\uparrow},c^{\dag}_{-k,\downarrow})$ is given by  
\begin{gather}
\mathcal{H}_{\rm BdG}(k) 
= \xi_k\tau_z + \alpha k \sigma_z \tau_0 + B\sigma_x \tau_z -\Delta\sigma_y\tau_y, 
\label{eq:Hbdg}
\end{gather}
where $\alpha$, $B$, and $\Delta$ are spin-orbit coupling, the Zeeman splitting energy, and the proximity-induced superconducting gap, respectively. The energy of free electrons with effective mass $m_{\rm eff}$ is given by $\xi_k\equiv k^2/2m_{\rm eff}-\mu$. We have introduced Pauli matrices ${\bm \sigma}$ and ${\bm \tau}$ in spin and particle-hole spaces, respectively. The quasiparticle energies are given by Eq.~\eqref{eq:Ek}.

We define the Green's functions in the Nambu space as
$G(k,z) = [z-\mathcal{H}_{\rm BdG}(k)]^{-1}$, where $z\in\mathbb{C}$ is the complex frequency. The $4\times 4$ matrix of the Green's function can be recast into
\begin{align}
G(k,z) = \begin{pmatrix}
G_0 + {\bm G}\cdot{\bm \sigma} & F_0i\sigma_y + i{\bm F}\cdot{\bm \sigma}\sigma_y \\ 
-i\sigma_y{F}_0 -i\sigma_y{\bm \sigma}\cdot\bar{\bm F} & \bar{G}_0 + \bar{\bm G}\cdot{\bm \sigma}^{\rm t}
\end{pmatrix}.
\label{eq:Gw0}
\end{align}
The expression of each component of the Green's functions can be derived by directly calculating the inverse matrix as 
\begin{align}
G_0(k,z) 
=& \frac{\xi_k(B^2-\Delta^2+z^2-\xi^2_k+\alpha^2k^2)}{D(k,z)} \nn \\
&-\frac{z(B^2+\Delta^2-z^2+\xi^2_k+\alpha^2k^2)}{D(k,z)}, \\
G_x(k,z)
=& \frac{B(-B^2+\Delta^2-\alpha^2k^2+(\xi_k+z)^2)}{D(k,z)},
\\
G_z(k,z) 
=&
\frac{\alpha k(-B^2-\Delta^2-\alpha^2k^2+(\xi_k+z)^2)}{D(k,z)},
\end{align}
and $G_y=0$. The off-diagonal components associated with pair amplitudes read 
\begin{align}
&F_0(k,z) = \frac{\Delta(z^2+B^2-\alpha^2k^2-\Delta^2-\xi^2_k)}{D(k,z)}, 
\label{eq:F0} \\
&{\bm F}(k,z)=\frac{2\Delta}{D(k,z)}(z B,i\alpha kB,\alpha k  \xi_k).
\label{eq:F}
\end{align}
The denominator is given by 
\begin{align}
D(k,z) 
=& \prod_{\sigma=\pm}[z-E_{\sigma}(k)][z+E_{\sigma}(-k)].
\label{eq:D}
\end{align}

\section{Density response in the Rashba superconductor}
\label{sec:wire}

In this section, we present the supplemental information on the renormalized density-density response functions, $\tilde{\chi}_{\rho\rho}$, defined in Eqs.~(3) and (4) in the main text. Here we choose the parameters, $m_{\rm eff}=0.014m_{\rm e}$, $\alpha = 0.5~{\rm eV}$\AA, $g=50$, and $\mu^{\rm s} = 1.0~{\rm meV}$~\cite{lut18}, which are the same as those in the main text.

\subsection{Denstiy response and gapless plasmons in the normal state} 
\label{sec:wiren}

In Fig.~\ref{fig:normal}, we plot the density-density response functions and collective excitation spectra in the normal state of the semiconducting wire with $\Delta=0$. The upper panels in Figs.~\ref{fig:normal}(b-e) show the imaginary part of the bare density-density response function, ${\rm Im}\chi^{\rm ww}_{11}(q,\omega)=-{\rm Im}\chi_{\rho\rho}(q,\omega)$, in the vicinity of the Lifsihtz transition ($\delta B=0$). For $\delta B<0$ ($B<B_{\rm L}$), two clear continua reflect the particle-hole excitations occurring in the inner and outer bands, which are labeled by the process I and II in Fig.~\ref{fig:normal}(a), respectively. The intraband particle-hole excitation process (III) is found to be small and invisible in the range of $q\le 0.01 {\rm nm}^{-1}$. At $\delta B=0$, the lowest continua vanish, signaling the Lifshitz point.

The lower panels in Figs.~\ref{fig:normal}(b-e) show the imaginary part of the charge fluctuation propagator in the semiconducting wire, $-{\rm Im}\mathcal{D}^{\rm ww}_{11}(q,\omega)$, {\it i.e.}, the excitation spectra of the collective modes (plasmons). The prominent gapless branch that exists throughout the entire range of $\delta B$ corresponds to the plasma excitation in the outer band. It is seen from the lower panel of Fig.~\ref{fig:normal}(c) that the additional gapless branch appears at $\delta B = -0.18~{\rm T}$ in between the continua I and II. This can be identified as the gapless plasmons arising from the collective excitations of electrons within the inner band. As the Lifshitz point is approached from $\delta B<0$, the lower branch exhibits softening and eventually disappears in the region of $\delta B>0$. The evolution of gapless plasmons and particle-hole continua around the Lifshitz point, as detailed in the main text, can be reflected by the renormalized density response functions.

\subsection{Denstiy response in the superconducting state} 
\label{sec:wires}

\begin{figure}[t!]
\includegraphics[width=\columnwidth]{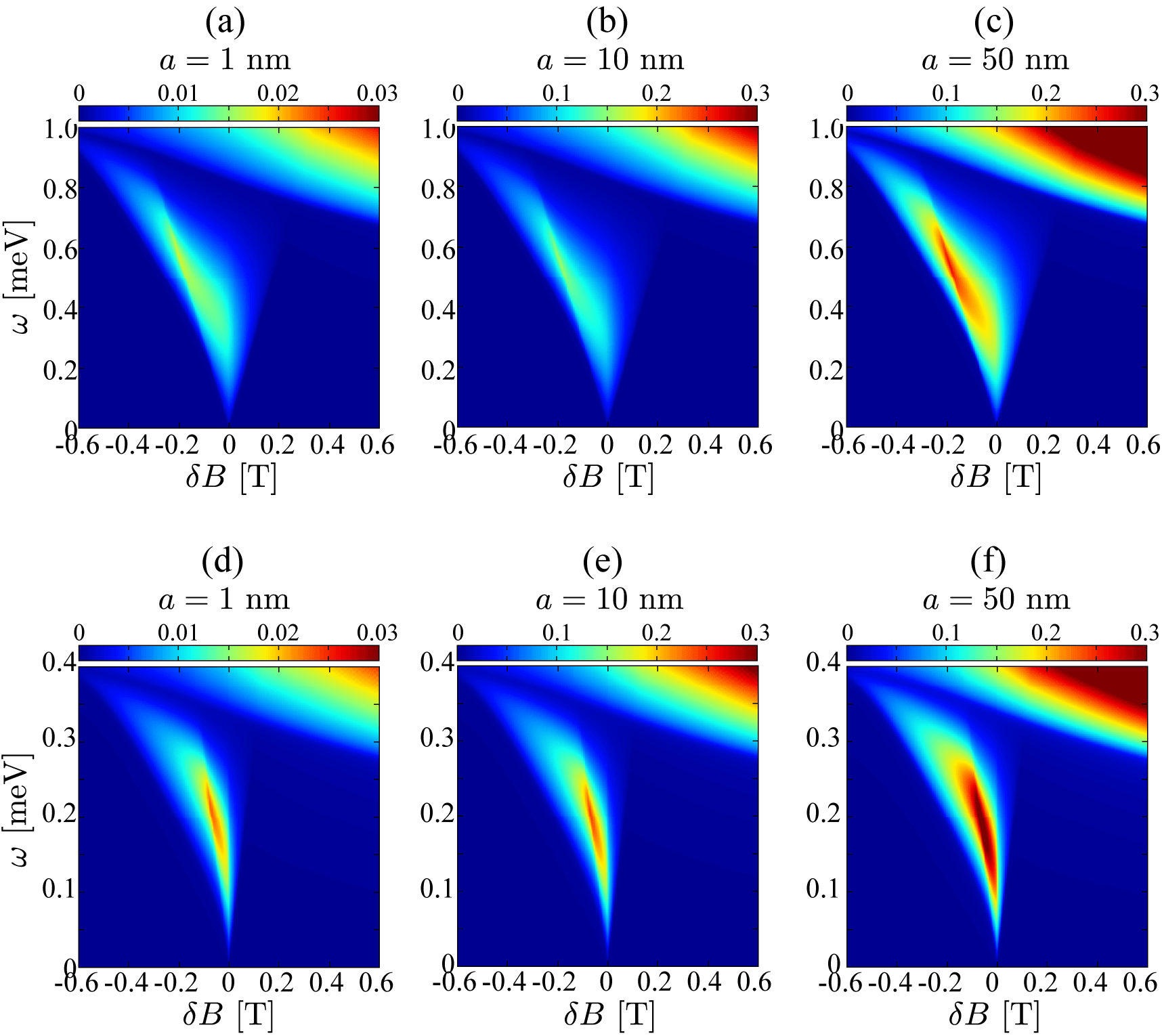}
\caption{Renormalized density response, $-{\rm Im}\tilde{\chi}_{\rho\rho}(q,\omega)$, as a function of $\delta B$ in the semiconducting wire with $\Delta = 0.5~{\rm meV}$ (a-c) and $0.2~{\rm meV}$ (d-f). The radius of the wire is set to $a=1~{\rm nm}$ (a,d), $10~{\rm nm}$ (b,e), and $50~{\rm nm}$ (c,f). In all data, we fix $q=1.0\times 10^{-4}~{\rm nm}^{-1}$.}
\label{fig:chiq0}  
\end{figure}

\begin{figure}[t!]
\includegraphics[width=\columnwidth]{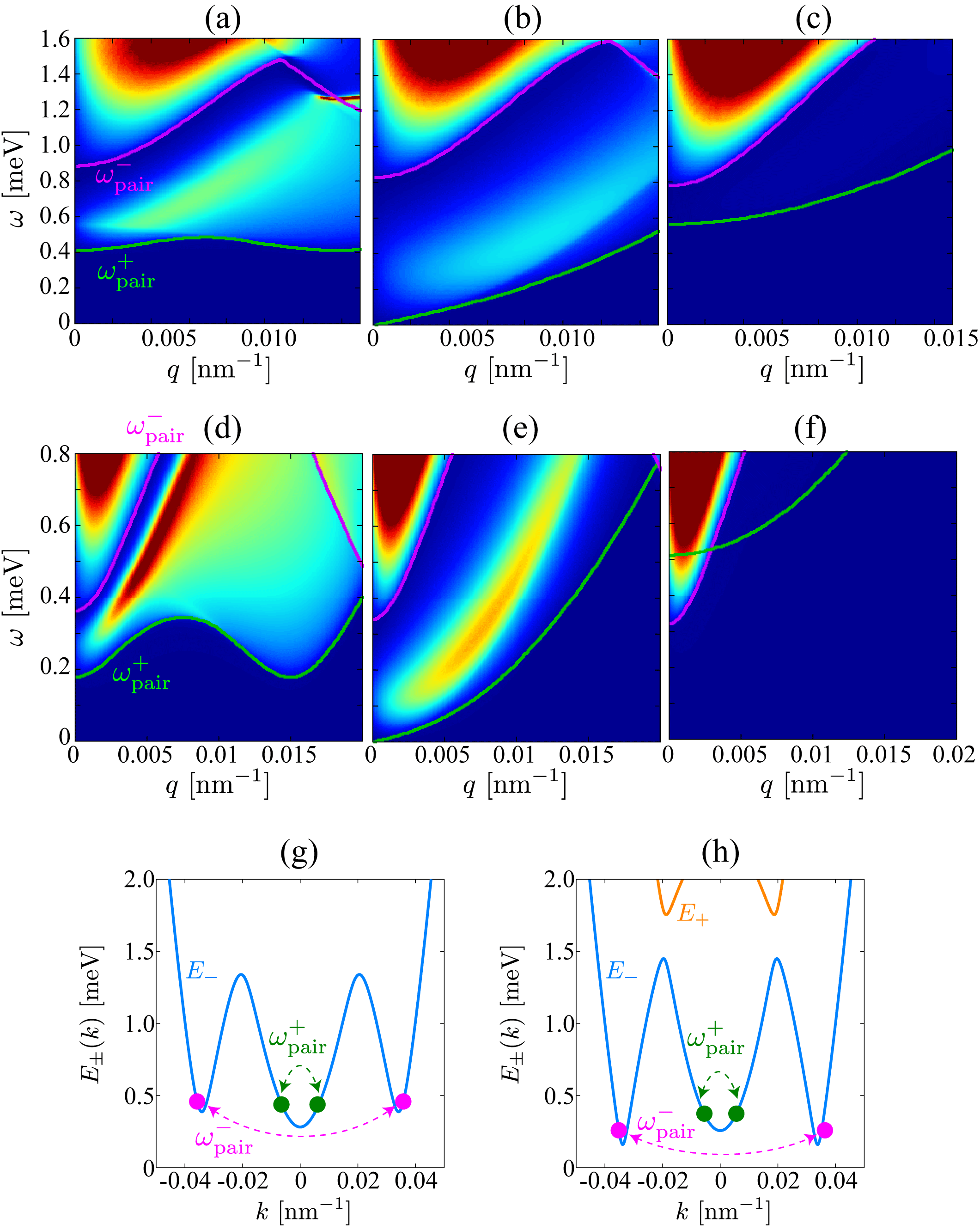}
\caption{(a-c,e-g) Renormalized density response, $-{\rm Im}\tilde{\chi}_{\rho\rho}(q,\omega)$, in the semiconducting wire with $\Delta = 0.5~{\rm meV}$ (a-c) and $0.2~{\rm meV}$ (d-f): $\delta B=-0.18~{\rm T}$ (a,d),  $\delta B=0$ (b,e), and $\delta B = 0.18~{\rm T}$ (c,f). In all data, we fix $a=100~{\rm nm}$. (g,h) Dispersion of bulk quasiparticle excitations at $\delta B = 0.18~{\rm T}$ with $\Delta = 0.5~{\rm meV}$ (g) and $0.2~{\rm meV}$ (h), where $\omega^+_{\rm pair}$ and $\omega^-_{\rm pair}$ denote the minimum pair excitation energy around the inner Fermi points $k=k_{{\rm F},+}^{\rm N}$ ($k=0$) and the outer Fermi points $k=k^{\rm N}_{{\rm F},-}$, respectively.}
\label{fig:chiq}  
\end{figure}

Let us discuss the density response of the Rashba superconductor for the different values of the proximitized gap $\Delta$. The normalized density response, $-{\rm Im}\tilde{\chi}_{\rho\rho}(q,\omega)$, which can be measured by electronic Raman spectroscopy, can capture the signal of the closing and reopening of the bulk excitation gap across the TPT. In Fig.~\ref{fig:chiq0}, we plot the field-dependence of the normalized density response for $\Delta=0.5~{\rm meV}$ and $0.2~{\rm meV}$. Figure~\ref{fig:chiq} also shows the dispersion of the normalized density response function in the trivial phase ($\delta B=-0.18~{\rm T}$), at the critical field ($\delta B =0$), and in the topological phase ($\delta B=0.18~{\rm T}$). We note that as $\Delta$ is the proximity-induced gap, its value is different from that of the superconducting gap in the parent superconductor, $\Delta^{\rm s}_0$, where the latter is determined by solving the gap equation in Eq.~\eqref{eq:gapeq}. It is seen from Figs.~\ref{fig:chiq0} and \ref{fig:chiq} that the qualitative features of the renormalized density response functions remain unaffected by changes in the value of $\Delta$ and the signal of the bulk gap closing and reopening can be observed through the evolution of $\tilde{\chi}_{\rho\rho}(q,\omega)$.

The sharp peaks observed in the normal state, which correspond to the gapless plasmons, disappear in the superconducting state, and the resulting density response functions in the trivial phase ($\delta B<0$) have two continua. As depicted in Figs.~\ref{fig:chiq}(d) and \ref{fig:chiq}(h), two continua are attributed to pair excitations in the inner and outer bands. The minimum pair excitation energies around the inner and outer Fermi points, $\omega^{\pm}_{\rm pair}$, coincide with the threshold of two continua. As shown in Figs.~\ref{fig:chiq}(b,c) and \ref{fig:chiq}(f,g), the bottom of the lower continuum touches zero at the TPT and becomes invisible in the topological phase.

In summary, the bulk gap closing and reopening across the TPT can be captured by the field evolution of the dynamical density-density response function. These noteworthy properties remain insensitive to the values of the proximitized gap $\Delta$.

\section{Collective modes and electronic Raman spectroscopy in hybrid systems}
\label{sec:CM}

In this section, we examine collective modes in semiconductor-superconductor hybrid systems and their contributions to dynamical density responses. In bulk superconductors, only the low-lying collective excitation is the Higgs mode and the phase mode is gapped out by the long-range Coulomb interaction through the Anderson-Higgs mechanism. Besides collective excitations, pair excitations occur in the low-energy region, with their threshold energy coinciding with that of the Higgs mode, $\omega = 2\Delta_{\rm s}$, where $\Delta_{\rm s}$ is the superconducting gap at equilibrium.

Here, as shown in Fig.~\ref{fig:setup0}(a), we consider a hybrid system in which a semiconducting wire is in contact with a conventional $s$-wave superconductor, with a Zeeman field $B$ applied along the wire. At equilibrium, the superconducting gap is proximitized to the wire, and the low-energy physics of electrons in the wire is well described by the Hamiltonian for Rashba superconductors in Eq.~\eqref{eq:Hbdg}. In this hybrid system, the low-lying excitations include the pair excitations in each subsystem, and plasmons in the semiconducting wire. As discussed in the main text, the field evolution of the pair excitation energies in the semiconducting wire reveals a bulk signature of the topological phase transition. We also take into account the tunneling processes of Copper pair fluctuations between the semiconducting wire and the superconductor and examine how their excitations contribute to the density response in hybrid systems. 

\begin{figure}[t!]
\includegraphics[width=\columnwidth]{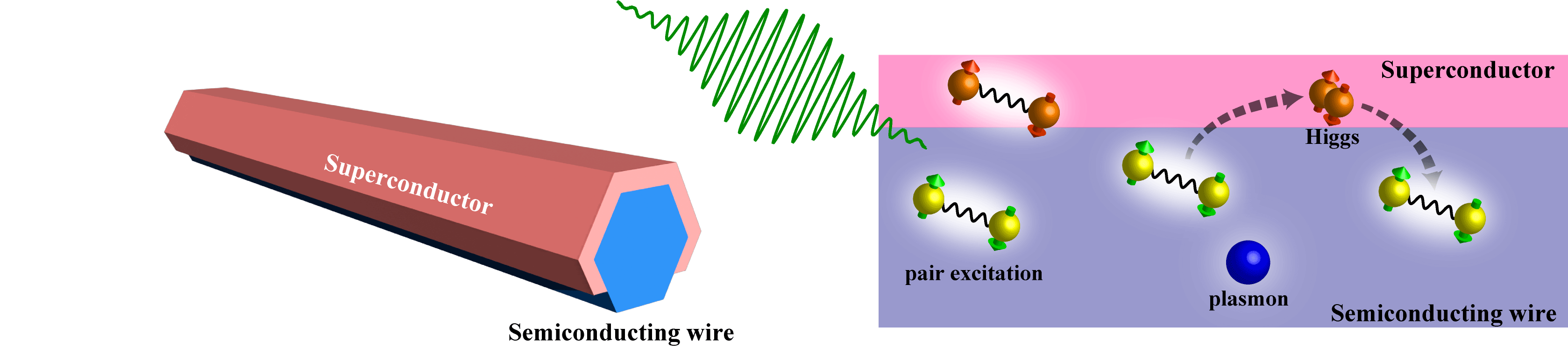}
\caption{(a) Schematics of our setup: A semiconducting wire is placed in contact with a $s$-wave conventional superconductor and a Zeeman field $B$ is applied along the wire. In realistic systems, the diameter of the wire is about $100~{\rm nm}$, while the thickness of the superconductor is $10~{\rm nm}$. As the latter scale is much longer than the Fermi wavelength, the superconductor is regarded as a three-dimensional bulk system. (b) Schematics of low-lying excitations in the hybrid system driven by external perturbation. The direct coupling of external fields with the Higgs mode is prohibited and the plasmons in the superconductor has an extremely large excitation gap ($\sim 1~{\rm eV}$).}
\label{fig:setup}  
\end{figure}

\subsection{Functional integral formalism}

Here, we explore collective modes within hybrid systems and how they influence dynamical density responses, using the functional integral formalism. The semiconducting wire comprises Rashba spin-orbit coupled electrons, while the superconductor consists of electrons that interact via a BCS-type pairing interaction. Two subsystems, the semiconductor and superconductor, are coupled through electron tunneling. The microscopic action then reads
\begin{widetext}
\begin{align}
\mathcal{S}[\bar{\psi}^{\rm w},\psi^{\rm w},\bar{\psi}^{\rm s},\psi^{\rm s}]
=& \int dx \bar{\psi}^{\rm w}_{\sigma}(x)
\left[ \partial_{\tau} + 
\varepsilon^{\rm w}_{\sigma\sigma^{\prime}}({\bm x})\right]
\psi^{\rm w}_{\sigma}(x)
+ \int dx\bar{\psi}^{\rm s}_{\sigma}(x)
\left[ \partial _{\tau} + \varepsilon^{\rm s}({\bm x})\right]
\psi^{\rm s}_{\sigma}(x) \nn \\
&-\int dx\Gamma(x) \left\{ 
\bar{\psi}^{\rm w}_{\sigma}(x)\psi^{\rm s}_{\sigma}(x) + {\rm h.c.}
\right\} 
-V^{\rm s}\int dx 
\bar{\psi}^{\rm s}_{\uparrow}(x) \bar{\psi}^{\rm s}_{\downarrow}(x)
\psi^{\rm s}_{\downarrow}(x) \psi^{\rm s}_{\uparrow}(x) \nn \\
&+\frac{1}{2}\int dx \int dx^{\prime}\delta n^{\rm w}(x)U(x-x^{\prime})\delta n^{\rm w}(x^{\prime}) 
+\frac{1}{2}\int dx \int dx^{\prime}\delta n^{\rm s}(x)U(x-x^{\prime})\delta n^{\rm s}(x^{\prime}) \nn \\
&+ \mathcal{S}_{\rm ext}[\bar{\psi}^{\rm w},\psi^{\rm w},\bar{\psi}^{\rm s},\psi^{\rm s},V^{\rm s},V^{\rm w}],
\label{eq:action}
\end{align}
where $\psi^{\rm w}_{\sigma}(x)$ and $\psi^{\rm s}_{\sigma}(x)$, are the electron operators with spin $\sigma=\uparrow,\downarrow$ in the semiconducting wire and the superconductor, respectively. We have also introduced the abbreviation, $x\equiv ({\bm x},\tau)$. In this SM, the variables with superscripts, ``${\rm w}$'' and ``${\rm s}$'', are defined in the semiconducting wire and the superconducting region, respectively. The first and second terms in Eq.~\eqref{eq:action} represent the Hamiltonian for normal electrons in the semiconducting wire and the superconductor, respectively.
The third term in Eq.~\eqref{eq:action} describes electron tunneling between two subsystems. The $s$-wave attractive interaction with the strength $V^{\rm s}>0$ is represented by the fourth term, which is absent in the semiconducting wire. 
The two terms in the third line of Eq.~\eqref{eq:action} correspond to the long-range Coulomb interaction between electrons in the semiconducting wire and in the superconductor, where $\delta n^{\rm w}(x)= \sum_{\sigma}\bar{\psi}^{\rm w}_{\sigma}(x)\psi^{\rm w}_{\sigma}(x) - n^{\rm w}_0$ and $\delta n^{\rm s}(x)= \sum_{\sigma}\bar{\psi}^{\rm s}_{\sigma}(x)\psi^{\rm s}_{\sigma}(x) - n^{\rm s}_0$ are the density fluctuation operators from the background charge density $n^{\rm w}_0$ and $n^{\rm s}_0$, respectively. The Coulomb interaction potential is defined as $U(x_1-x_2)=U({\bm x}_{12})\delta(\tau_{12})$, where 
$U({\bm x}_{12})=\frac{e^2}{|{\bm x}_{12}|}$, with ${\bm x}_{ij}\equiv {\bm x}_i-{\bm x}_j$.
The last term in Eq.~\eqref{eq:action} describes the coupling of electron densities in the semiconducting wire and the superconductor with external perturbations, $V^{\rm s}$ and $V^{\rm w}$. The action is given by 
\beq
\mathcal{S}_{\rm ext} = \sum_{\sigma} \int dx \left[
V^{\rm s}(x)\bar{\psi}^{\rm s}_{\sigma}(x)\psi^{\rm s}_{\sigma}(x)
+V^{\rm w}(x)\bar{\psi}^{\rm w}_{\sigma}(x)\psi^{\rm w}_{\sigma}(x) 
\right].
\eeq

We take the $x$-axis along the semiconducting wire and set the center of the wire at ${\bm x}_{\perp}\equiv (y,z) ={\bm 0}$. We assume that the radius $a$ of the semiconducting wire is small enough for all the electrons to occupy the lowest subband of transverse motion of $\varepsilon^{\rm w}({\bm x})$. Thus the electron field in the semiconducting wire takes the form 
\beq
\psi^{\rm w}(x) = \varphi_0({\bm x}_{\perp})\psi^{\rm w}(x_1), \quad 
 \varphi_0({\bm x}_{\perp})=\sqrt{\frac{2}{\pi a^2}}e^{-x^2_{\perp}/a^2},
\eeq
where $x_1 \equiv (x,\tau)$. The microscopic action in Eq.~\eqref{eq:action} is then recast into 
\begin{align}
\mathcal{S}[\bar{\psi}^{\rm w},\psi^{\rm w},\bar{\psi}^{\rm s},\psi^{\rm s}]
=& \int dx_1 \bar{\psi}^{\rm w}_{\sigma}(x_1)
\left[\partial_{\tau}+\varepsilon^{\rm w}_{\sigma\sigma^{\prime}}(x_1)\right]
\psi^{\rm w}_{\sigma}(x_1)
+ \int dx\bar{\psi}^{\rm s}_{\sigma}(x)
\left[\partial_{\tau}+\varepsilon^{\rm s}({\bm x})\right]
\psi^{\rm s}_{\sigma}(x) \nn \\
& -\Gamma\int dx_1 \left\{ 
\bar{\psi}^{\rm w}_{\sigma}(x_1)\psi^{\rm s}_{\sigma}(x_1,{\bm x}_{\perp}={\bm 0}) + {\rm h.c.}
\right\} 
-V^{\rm s}\int dx 
\bar{\psi}^{\rm s}_{\uparrow}(x) \bar{\psi}^{\rm s}_{\downarrow}(x)
\psi^{\rm s}_{\downarrow}(x) \psi^{\rm s}_{\uparrow}(x) \nn \\
&+\frac{1}{2}\int dx_1 \int dx^{\prime}_1\delta n^{\rm w}(x_1)U^{\rm w}(x_1-x^{\prime}_1)\delta n^{\rm w}(x^{\prime}_1) 
+\frac{1}{2}\int dx \int dx^{\prime}\delta n^{\rm s}(x)U^{\rm s}(x-x^{\prime})\delta n^{\rm s}(x^{\prime}) \nn \\
&+ \mathcal{S}_{\rm ext}[\bar{\psi}^{\rm w},\psi^{\rm w},\bar{\psi}^{\rm s},\psi^{\rm s},V^{\rm s},V^{\rm w}],
\label{eq:action2}
\end{align}
The single-particle energies are given in the momentum representation as
\begin{gather}
\varepsilon^{\rm w}(k_x) = \frac{k^2_x}{2m_{\rm eff}} + \alpha k_x\sigma_z + B\sigma_x-\mu^{\rm w}, 
\label{eq:epsilon_wire}\\
\varepsilon^{\rm s}({\bm k}) = \frac{{\bm k}^2}{2m_{\rm s}}-\mu^{\rm s},
\end{gather}
where and $m_{\rm s}$ is the effective mass of electrons in the parent superconductor. 
Electron tunneling between two subsystems is given by $\Gamma(x)\equiv \Gamma \delta({\bm x}_{\perp})$.
The Fourier transform of the Coulomb potential in a cylindrical wire of radius $a$ is given by 
\beq
U^{\rm w}(q) = -\frac{e^2}{4\pi}e^{q^2a^2} {\rm Ei}\left( -q^2a^2 \right),
\label{eq:Uq}
\eeq
where ${\rm Ei}(x) = -\int ^{\infty}_{-x} \frac{e^u}{u}du$ is the exponential-integral function~\cite{cayao_phd,giu}. The Coulomb potential in three dimensional superconductors is $U^{\rm s}(q)=4\pi e^2/q^2$.

In a conventional superconductor, {\it e.g.}, Al, the chemical potential and the superconducting gap are $\mu^{\rm s}=O(1~{\rm eV})$ and $\Delta^{\rm s}=O(0.1~{\rm meV})$, 
respectively. The typical thickness of the superconductor is about $10~{\rm nm}$ in realistic situations~\cite{microsoft}, which is much larger than the Fermi wavelength of electrons in the superconductor, $k_{\rm F}^{\rm s-1}=\sqrt{2m_{\rm s}\mu^{\rm s}/\hbar^2}\sim 1~\mbox{\AA}$, where we take $m_{\rm s}=m_{\rm e}$ and $\mu^{\rm s}=10~{\rm eV}$. Therefore, below, we consider the superconductor as a three-dimensional bulk system.

We then perform the Hubbard-Stratonovich transformation in Eq.~\eqref{eq:action2}, introducing the auxiliary bosonic fields $\Delta^{\rm s}$, $\phi^{\rm s}$, and $\phi^{\rm w}$, where $\Delta^{\rm s}$ is the complex field representing the superconducting order, and $\phi^{\rm s}$ and $\phi^{\rm w}$ are the internal electric potential produced by electrons in the superconductor and the semiconductor, respectively. The resulting action reads
\begin{align}
{\mathcal{S}}[\bar{\psi}^{{\rm w}/{\rm s}},\psi^{{\rm w}/{\rm s}},\bar{\Delta}^{\rm s},\Delta^{\rm s},\phi^{{\rm w}/{\rm s}}] =& -\frac{1}{2}\sum_{k,q}\bar{\Psi}_{k+q/2}
\left\{ 
-i\varepsilon_n + \begin{pmatrix}
\mathcal{H}^{\rm w}(k_x,q_1; V^{\rm w})\delta({\bm k}_{\perp}) & \Gamma \tau_z\\ 
\Gamma\tau_z & \mathcal{H}_{\rm sc}({\bm k},q; V^{\rm s})
\end{pmatrix}
\right\}
\Psi_{k-q/2}\nn \\
&+ \sum _q \frac{|\Delta^{\rm s}(q)|^2}{V^{\rm s}}
+e^2\sum_q\frac{\phi^{\rm s}(q)\phi^{\rm s}(-q)}{2U^{\rm s}_q}
+e^2\sum_q\frac{\phi^{\rm w}(q)\phi^{\rm w}(-q)}{2U^{\rm w}_q} ,
\label{eq:action3}
\end{align}
where $\Psi_k=(
\psi^{\rm w}_{k_1,\uparrow},\psi^{\rm w}_{k_1,\downarrow},
\bar{\psi}^{\rm w}_{-k_1,\uparrow},\bar{\psi}^{\rm w}_{-k_1,\downarrow},
\psi^{\rm s}_{k,\uparrow},\psi^{\rm s}_{k,\downarrow},
\bar{\psi}^{\rm s}_{-k,\uparrow},\bar{\psi}^{\rm s}_{-k,\downarrow})$. Here we have introduced, $k_1\equiv (i\varepsilon_n,k_x)$, $q_1\equiv (i\omega_m,q_x)$, $k\equiv (i\varepsilon_n,{\bm k})$, and $q\equiv(i\omega_m,{\bm q})$, where $\varepsilon_n=(2n+1)\pi T$ and $\omega_m=2m \pi T$ are the Matsubara frequencies for fermions and bosons at temperature $T$, respectively ($n,m\in\mathbb{Z}$). ${\bm k}_{\perp}\equiv (k_y,k_z)$ is the momentum perpendicular to the semiconducting wire. We have also introduced the shorthand notation, $\sum _k \equiv \int \frac{d{\bm k}}{(2\pi)^3} T\sum_n $. The Hamiltonian for the semiconducting wire is diagonal in the particle-hole space as 
\beq
\mathcal{H}^{\rm w}(k_x,q_1;V^{\rm w}) = \begin{pmatrix}
\varepsilon^{\rm w}(k_x) & 0 \\
0 & -\varepsilon^{\rm t}_{\rm w}(-k_x)
\end{pmatrix}\delta_{q_1,0}
+ie\phi^{\rm w}(q_1)\tau_z + V^{\rm w}(q_1)\tau_z,
\eeq
including the coupling term of electrons in the wire with an external potential, $V^{\rm w}(q_1)=V^{\rm w}(q_1,{\bm q}_{\perp}={\bm 0})$. Here the superscript ``t'' indicates the transpose of a matrix. The Hamiltonian in the superconducting region reduces to the Bogoliubov-de Gennes Hamiltonian as 
\beq
\mathcal{H}^{\rm s}({\bm k},q; V^{\rm s}) = \begin{pmatrix}
\varepsilon^{\rm s}({\bm k})\delta_{q,0} +ie\phi^{\rm s}(q) + V^{\rm s}(q) & \Delta^{\rm s}(q) \\
\Delta^{{\rm s}\ast}(q) & -\varepsilon^{\rm s}({\bm k})\delta_{q,0}
-ie\phi^{\rm s}(q)- V^{\rm s}(q)
\end{pmatrix} .
\eeq

Let us now discretize the two-dimensional momentum planes $(k_y, k_z)$ and $(q_y,q_z)$ into $N_k$ and $N_q$ grids with an infinitesimal spacing $d$, respectively, and introduce the $(4+4N_kN_q)$-dimensional field, 
\beq
{\bm \Psi}_{k_1} \equiv \left[
\psi^{\rm w}_{k_1,\uparrow},
\psi^{\rm w}_{k_1,\downarrow},
\bar{\psi}^{\rm w}_{-k_1,\uparrow},
\bar{\psi}^{\rm w}_{-k_1,\downarrow},
\left({\bm \psi}^{\rm s}_{k_1,\uparrow}\right)^{\rm t},
\left({\bm \psi}^{\rm s}_{k_1,\downarrow}\right)^{\rm t},
\left(\bar{\bm \psi}^{\rm s}_{-k_1,\uparrow}\right)^{\rm t},
\left(\bar{\bm \psi}^{\rm s}_{-k_1,\downarrow}\right)^{\rm t}\right]^{\rm t}.
\eeq
The $N_kN_q$-dimensional vector fields, ${\psi}^{\rm s}_{k_1,ij,\sigma}\equiv [{\bm \psi}^{\rm s}_{k_1,\sigma}]_{ij}$, represents the electron fields defined on the grids in $\forall {\bm k}_{\perp}$ and and $\forall {\bm q}_{\perp}$, where $i$ ($j$) denotes the label of the grids in the ${\bm k}_{\perp}$ (${\bm q}_{\perp}$) plane. Then, we express the first term in Eq.~\eqref{eq:action2} in terms of the matrix form as
\begin{align}
-\frac{1}{2}\sum_{k_1,q_1} \bar{\bm \Psi}_{k_1+q_1/2}
\begin{pmatrix}
-i\varepsilon_n+\mathcal{H}^{\rm w}(k_x,q_1) & {\bf T} \\
{\bf T}^{\rm t} & 
-i\varepsilon_n{\bf I} + {\bf H}^{\rm s}(k_x,q_1)
\end{pmatrix}
{\bm \Psi}_{k_1-q_1/2}
\equiv \frac{1}{2}\sum_{k_1,q_1} \bar{\bm \Psi}_{k_1+q_1/2}
{\bf G}^{-1}(k_1,q_1)
{\bm \Psi}_{k_1-q_1/2},
\label{eq:matrix}
\end{align}
We have introduced the $(4+4N_kN_q)\times (4+4N_kN_q)$ matrix form of the full Green's function in hybrid systems as 
\beq
{\bf G}(k_1,q_1)
= 
\begin{pmatrix}
i\varepsilon_n-\mathcal{H}^{\rm w}(k_x,q_1) & -{\bf T} \\
-{\bf T}^{\rm t} & 
i\varepsilon_n{\bf I} - {\bf H}^{\rm s}(k_x,q_1)
\end{pmatrix}^{-1}
\eeq
where 
${\bf I}$ is the $4N_kN_q\times 4N_kN_q$ unit matrix and the $4N_kN_q\times 4N_kN_q$ matrix, ${\bf H}^{\rm s}$, is defined as 
\begin{align}
\left[{\bf H}^{\rm s}(k_x,q_1)\right]_{ij,i^{\prime}j^{\prime}} 
\equiv \delta_{i,i^{\prime}}\delta_{j,j^{\prime}}\left(\frac{d}{2\pi}\right)^4 \mathcal{H}^{\rm s}(k_x,{\bm k}^i_{\perp},q_x,{\bm q}^j_{\perp},i\omega_m).
\label{eq:Hmat}
\end{align}
The electron tunneling is represented in the $4\times 4N_kN_q$ matrix as
\begin{align}
\left[{\bf T}\right]_{1,i^{\prime}j^{\prime}} 
\equiv \left(\frac{d}{2\pi}\right)^4\Gamma \tau_z.
\end{align}
By integrating out the fermion fields ${\bm \Psi}$ and $\bar{\bm \Psi}$, the effective action in Eq.~\eqref{eq:action2} can be recast into  
\begin{align}
{\mathcal{S}}[\phi^{\rm w},\Delta^{\rm s},\phi^{\rm s},V^{\rm w},V^{\rm s}] =& -\frac{1}{2}{\rm Tr}\log [-{\bf G}^{-1}(k_1,q_1)]
+ \sum _q \frac{|\Delta^{\rm s}(q)|^2}{V^{\rm s}}
+e^2\sum_q\frac{\phi^{\rm s}(q)\phi^{\rm s}(-q)}{2U^{\rm s}_q}
+e^2\sum_q\frac{\phi^{\rm w}(q)\phi^{\rm w}(-q)}{2U^{\rm w}_q} .
\label{eq:action4}
\end{align}
\end{widetext}

\subsection{Effective action for the semiconducting wire in equilibrium}

Let $\Delta^{\rm s}_0$ and $\phi^{\rm s/w}_0$ be the saddle-point values of the superconducting order and the Coulomb potential, respectively. 
We impose the charge neutrality condition that electron charges are balanced by background positive charges in the equilibrium, resulting in $\phi^{\rm s/w}_0=0$. Below, we derive the equilibrium Green's functions and the gap equation that self-consistently determines $\Delta^{\rm s}_0$. 

Let $G_0^{{\rm w}}(k_1)$ and ${\bf G}^{\rm s}_0(k_1)$ be the bare Green's functions in the semiconducting wire and the superconductor at equilibrium, respectively, which are defined as
\begin{gather}
G_0^{{\rm w}}(k_1)\equiv 
\left[i\varepsilon _n - 
\begin{pmatrix}
\varepsilon_{\rm w}(k) & 0 \\
0 & -\varepsilon^{\rm t}_{\rm w}(-k)
\end{pmatrix}
\right]^{-1}, \\
{\bf G}_0^{{\rm s}}(k_1)\equiv 
\left[i\varepsilon _n{\bf I} -
{\bf H}_{0}^{\rm s}(k_1)\right]^{-1} ,
\end{gather}
where ${\bf H}_{0}^{\rm s}(k_1)$ is the $4N_k\times 4N_k$ matrix form of the BdG Hamiltonian obtained from Eq.~\eqref{eq:Hmat} with replacing $\Delta^{\rm s}(q)$ with $\Delta^{\rm s}_0\in\mathbb{R}$ and ignoring $\phi^{\rm s}$ and $V^{\rm s}$. The bare Green's function in the superconducting region, $G^{\rm s}_0(k)=G^{\rm s}_0(k_1,{\bm k}^{i}_{\perp})=[{\bf G}^{\rm s}_0(k_1)]_{i,i}$, is given by 
\begin{align}
G^{\rm s}_0(k) = -\frac{i\varepsilon_n + \varepsilon^{\rm s}({\bm k})\tau_z + \Delta^{\rm s}_0\tau_x}{\varepsilon^2_n +[E^{\rm s}_0({\bm k})]^2},
\label{eq:Gs0}
\end{align}
where $E^{\rm s}_0({\bm k})= \sqrt{[\varepsilon^{\rm s}({\bm k})]^2+\Delta^{{\rm s}}_0}$ is the quasiparticle excitation energy in the conventional $s$-wave superconductor.
The saddle-point action is then obtained from Eq.~\eqref{eq:action3} as
\begin{align}
\mathcal{S}_0[\Delta^{\rm s}_0] 
= -\frac{1}{2}{\rm Tr} \log \left[-{\bf G}^{-1}_0(k_1)\right]
+ \sum _q \frac{|\Delta^{\rm s}_0|^2}{V_{\rm s}},
\end{align}
where ${\bf G}_0(k)$ is the $4(N_k+1)\times 4(N_k+1)$ matrix of the equilibrium Green's function 
\begin{align}
{\bf G}_0(k_1) =& \begin{pmatrix}
G^{{\rm w}-1}_0(k_1) & -{\bf T} \\ 
-{\bf T}^{\rm t} & {\bf G}^{{\rm s}-1}_0(k_1)
\end{pmatrix}^{-1} \nn \\
\equiv& \begin{pmatrix}
G^{\rm w}(k_1) & {\bf G}^{\rm ws}(k_1) \\ 
{\bf G}^{\rm sw}(k) & {\bf G}^{\rm s}(k_1)
\end{pmatrix}.
\end{align}

\begin{figure}[t!]
\includegraphics[width=\columnwidth]{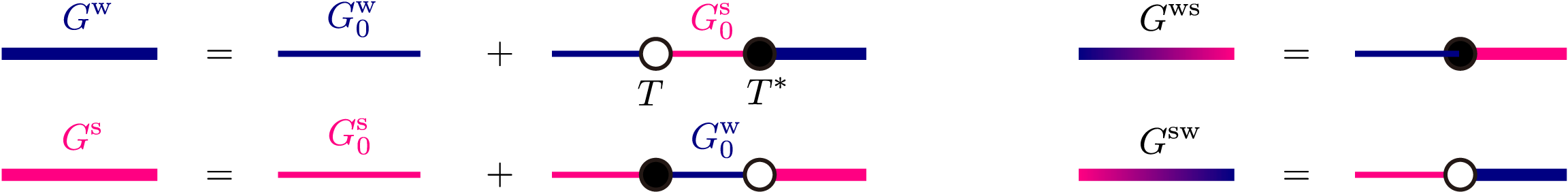}
\caption{Diagrammatic representations of the renormalized Green's functions, $G^{\rm w}$, $G^{\rm w}$, $G^{\rm ws}$, and $G^{\rm sw}$, where $G^{\rm w}$ ($G^{\rm s}$) is the propagator in the semiconductor (superconductor) and $G^{\rm ws}$ and $G^{\rm sw}$ are the propagators of quasiparticles tunneling between two subsystems. The thin and thick lines represent the bare and normalized Green's functions, respectively, and the open and filled circles denote the tunneling $T=T^{\ast}=\Gamma\tau_z$.}
\label{fig:diagram}  
\end{figure}

The diagonal components, $G^{\rm w}(k_1)$ and $G^{\rm s}(k)=G^{\rm s}(k_1,{\bm k}^i_{\perp})\equiv [{\bf G}^{\rm s}(k_1)]_{i,i}$, are the renormalized propagators in the semiconducting wire and the superconductor, respectively, while the off-diagonal components, $G^{\rm ws}$ and $G^{\rm sw}$, represent the propagators across two subsystems. The diagrammatic representations are depicted in Fig.~\ref{fig:diagram}(a). The diagonal components obey the Dyson equations,
\begin{gather}
G^{\rm w}(k_1) = G^{\rm w}_0(k_1) + G^{\rm w}_0(k_1)\Sigma^{\rm w}_0(k_1) G^{\rm w}(k_1), \\ 
\Sigma^{\rm w}_0(k_1) = \Gamma^2\int \frac{d{\bm k}_{\perp}}{(2\pi)^2} \tau_z G^{\rm s}_0({\bm k},i\varepsilon_n)\tau_z,
\label{eq:Gw}
\end{gather}
and 
\begin{gather}
G^{\rm s}(k) = G^{\rm s}_0(k) +  G^{\rm s}_0(k)\Sigma^{\rm s}_0(k_1) G^{\rm s}(k), \\ 
\Sigma^{\rm s}_0(k_1) = \Gamma^2 \tau_z G^{\rm w}_0(k_x,i\varepsilon_n)\tau_z.
\label{eq:Gs}
\end{gather}
The off-diagonal components are determined by
\begin{gather}
G^{\rm ws}(k) =G^{\rm w}_0(k_1) \Gamma \tau_z G^{\rm s} (k), \\
G^{\rm sw}(k) =G^{\rm s}_0(k) \Gamma \tau_z G^{\rm w} (k_1).
\end{gather}

The effective action for electrons in the semiconducting wire can be also obtained from Eq.~\eqref{eq:action3} by integrating out the electron field in the superconducting region ($\psi^{\rm w}_{\bm k}$).
\beq
{\mathcal{S}}^{\rm w}_0[\bar{\psi}^{\rm w},\psi^{\rm w}] 
=
\frac{1}{2}\sum_{k_1}
\bar{\psi}^{\rm w}_{k_1}
\left[G^{{\rm w}}(k_1)\right]^{-1}
\psi^{\rm w}_{k_1} + \frac{|\Delta^{\rm s}_0|^2}{V^{\rm s}}.
\label{eq:Swire}
\eeq
The renormalized Green's function for the semiconducting wire, $G^{\rm w}$, obeys the Dyson equation \eqref{eq:Gw}. The off-diagonal components of $\Sigma^{\rm w}_0(k_1)$ in the particle-hole space describe the tunneling of Cooper pairs from the superconducting region, while the diagonal components shift the chemical potential weakly depending on frequency and momentum. As the latter is not essential, we ignore the renormalization effect of the chemical potential. As a result, the renormalized Green's function in Eq.~\eqref{eq:Gw} reduces to Eq.~\eqref{eq:Gw0}.

In the same manner, the effective action for electrons in the superconductor is obtained by integrating out the electron field in the semiconducting wire ($\psi^{\rm w}_k$). The renormalized Green's function in the superconductor is given by Eq.~\eqref{eq:Gs}. The self-energy, $\Sigma^{\rm s}_0(k)$, describes electron tunneling from the semiconducting wire, leading to the renormalization of the chemical potential as $\mu_{\rm s}\tau_z\rightarrow \mu_{\rm s}\tau_z+T^{\ast}G^{\rm w}(k)T$. As $\mu_{\rm sc}$ is about $O(1~{\rm eV})$ in real materials, however, the renormalization effect is negligible and the renormalized Green's function reduces to the bare Green's function in Eq.~\eqref{eq:Gs0}, {\it i.e.}, $G^{\rm s}(k)\approx G^{\rm s}_0(k)$. The effective action then reads
\beq
{\mathcal{S}}^{\rm s}_0[\bar{\psi}^{\rm s},\psi^{\rm s},\Delta^{\rm s}_0] 
= \frac{1}{2}\sum_{k}
\bar{\psi}^{\rm s}_{k}
\left[G^{{\rm s}}_0(k)\right]^{-1}
\psi^{\rm s}_k + \frac{|\Delta^{\rm s}_0|^2}{V^{\rm s}}.
\eeq
Integrating out the fermion fields $\psi^{\rm s}$ and $\bar{\psi}^{\rm s}$, the effective action is recast into
\begin{align}
{\mathcal{S}}^{\rm s}_0[\Delta^{\rm s}_0] 
= -\frac{1}{2}{\rm Tr} \log \left(-G^{{\rm s}-1}_0\right)
+ \sum _q \frac{|\Delta^{\rm s}_0|^2}{V^{\rm s}}.
\end{align}
The equilibrium pair potential is obtained from the stationary condition of the saddle-point action, $\delta \mathcal{S}^{\rm s}_0/\delta \Delta^{\rm s}_0=0$. This reduces to the gap equation for conventional $s$-wave superconductors, 
\beq
\frac{1}{V^{\rm s}} = \int \frac{d^3{\bm k}}{(2\pi)^3}\frac{\tanh(E^{\rm s}_{\bm k}/2T)}{2E^{\rm s}_{\bm k}}.
\label{eq:gapeq}
\eeq

\subsection{Fluctuation action for the semiconducting wire}

Collective modes in superconductors comprise the amplitude and phase fluctuations of the superconducting order. We define the space-time fluctuations of the $s$-wave superconducting order around $\Delta^{\rm sc}_0$ as $\delta \Delta^{\rm s}_q = \Delta^{\rm s}_q - \Delta^{\rm s}_0$. Then, the self-energy fluctuation in the superconducting region is given by
\begin{align}
\delta\Sigma^{\rm s}(q)
= \frac{1}{2}i\sigma_y\left[i\tau_y\delta\Delta^{{\rm s}+}(q) + \tau_x\delta\Delta^{{\rm s}-}(q)\right] + ie\phi^{\rm s}(q)\tau_z,
\end{align}
where $\delta \Delta^{{\rm s}+}(q)\equiv \delta\Delta^{\rm s}(q) + \delta\Delta^{{\rm s}\ast}(-q)$ and $\delta \Delta^{{\rm s}-}(q)\equiv \delta\Delta^{\rm s}(q) - \delta\Delta^{{\rm s}\ast}(-q)$ correspond to the fluctuations of the real part (amplitude) and imaginary part (phase) of the gap function, respectively.

\begin{widetext}
To clarify the contributions of the collective modes in the superconductor to dynamical density responses, we here derive the fluctuation action for semiconductor-superconductor hybrid systems. Here we take the ${\bm q}$ vector along the direction of the semiconducting wire, {\it i.e.}, ${\bm q}=q\hat{\bm x}$. We start with the action in Eq.~\eqref{eq:action4}
\begin{align}
\mathcal{S}[\Delta^{\rm s},\phi^{{\rm w}/{\rm s}},V^{{\rm w}/{\rm s}}] 
= -\frac{1}{2}{\rm Tr}\log \left[ 
-{\bf G}^{-1}\right] 
+ \sum _q \frac{|\Delta^{\rm s}(q)|^2}{V_{\rm s}}
+e^2\sum_q\frac{\phi^{\rm s}(q)\phi^{\rm s}(-q)}{2U^{\rm s}_q}
+e^2\sum_{q_1}\frac{\phi^{\rm w}(q_1)\phi^{\rm w}(-q_1)}{2U^{\rm s}_{q_1}}.
\label{eq:sfluct}
\end{align}
The inverse matrix of the full Green's function is defined with the equilibrium Green's function ${G}_0$ as 
\begin{align}
{\bf G}^{-1}(k,q)
= {\bf G}^{-1}_0(k) - \delta{\bf \Sigma}(q), 
\label{eq:Ginv}
\end{align}

The linear fluctuation of the self-energy consists of the fluctuations of the superconducting gap and Coulomb potentials and the external field coupled to electron densities as
\begin{align}
\delta {\bf \Sigma}(q) 
=& \frac{1}{2}
\begin{pmatrix}
2ie\phi^{\rm w}(q)\tau_z  & 0 \\
0 & -\sigma_y\tau_y\delta{ \Delta}^{{\rm s}+}(q)
+ i\sigma_y\tau_x\delta{\Delta}^{{\rm s}-}(q) + 2ie{\bf \phi}^{\rm s}(q)\tau_z 
\end{pmatrix} 
+ \begin{pmatrix}
V^{\rm w}(q)\tau_z & 0 \\
0 & V^{\rm s}(q)\tau_z
\end{pmatrix} .
\label{eq:sigma_exp}
\end{align}
Substituting Eqs.~\eqref{eq:Ginv} and \eqref{eq:sigma_exp} into Eq.~\eqref{eq:sfluct}, the fluctuation action, $\mathcal{S}_{\rm fluc} \equiv \mathcal{S}-\mathcal{S}_0$, reads
\begin{align}
\mathcal{S}_{\rm fluc}
= \frac{1}{2}\left[\frac{1}{2}{\rm Tr}\left({\bf G}_0\delta {\bf \Sigma}\right)^2 + \sum _q \sum_{c=\pm}\frac{c\delta\Delta^{{\rm s}c}_q\delta\Delta^{{\rm s}c}_{-q}}{2V_{\rm s}}
+e^2\sum_q\frac{\delta\phi^{\rm s}_q\delta\phi^{\rm s}_{-q}}{U^{\rm s}_q}
+e^2\sum_{q_1}\frac{\delta\phi^{\rm w}_{q_1}\delta\phi^{\rm w}_{-{q_1}}}{U^{\rm w}_{q_1}}\right].
\label{eq:action_fluc}
\end{align}
The Gaussian fluctuation approximation incorporates the coupling of fluctuations of superconducting order to the long-range Coulomb potential and describes all branches of the collective excitations. The first term in Eq.~\eqref{eq:action_fluc} consists of the bilinear form of fluctuations in two subsystems,
\begin{align}
\frac{1}{2}{\rm Tr}\left({\bf G}_0\delta {\bf \Sigma}\right)^2
=& \sum_{a,b={\rm s},{\rm w}}\sum_q
\left[
\delta\Sigma^{a}_i(q) \chi^{ab}_{ij}(q) \delta\Sigma^{b}_j(-q)
+\delta\Sigma^{a}_i(q) \chi^{ab}_{iV}(q) V^{b}(-q) \right. \nn \\
& \left. 
+V^a(q) \chi^{ab}_{Vj}(q) \delta\Sigma^{b}_j(-q)
+V^{a}(q) \chi^{ab}_{VV}(q) V^{b}(-q)
\right],
\label{eq:tt}
\end{align}
where the self-energy fluctuations are given by $\delta \Sigma^{\rm w}_{i=1}\equiv \delta\phi^{\rm w}$ and $(\delta \Sigma^{\rm s}_1,\delta \Sigma^{\rm s}_2,\delta \Sigma^{\rm s}_3)\equiv (\delta\Delta^{{\rm s}+},\delta\Delta^{{\rm s}-},\delta\phi^{\rm s})$.
The density-density correlation functions in the semiconducting wire and the superconducting region are defined as 
\begin{align}
&\chi^{\rm ww}_{ij}(q) = \frac{1}{2}\sum _{k_1} {\rm tr}_4
\left[
G^{\rm w}(k_1)\Lambda^{\rm w}_i G^{\rm w}(k_1-q)\Lambda^{\rm w} _j
\right], 
\label{eq:chiww} \\
&\chi^{\rm ss}_{ij}(q) = \frac{1}{2}\sum _k {\rm tr}_4
\left[
G^{\rm s}(k)\Lambda^{\rm s}_i G^{\rm s}(k-q)\Lambda^{\rm s}_j 
\right], \label{eq:chiss} 
\end{align}
respectively, where the vertex functions are 
$\Lambda^{\rm w}_{1}=ie\tau_z$, $\Lambda^{\rm s}_1 = -\frac{1}{2}\sigma_y\tau_y$, $\Lambda^{\rm s}_2=\frac{1}{2}i\sigma_y\tau_x$, and $\Lambda^{\rm s}_3 =ie\tau_z$. The correlation functions in conventional superconductors obey $\chi^{\rm ss}_{ij}(q) = \chi^{\rm ss}_{ji}(q)$ as $E_k=E_{-k}$. 
The Green's functions, $G^{\rm w}$ and $G^{\rm s}$, obey the Dyson equation in Eqs.~\eqref{eq:Gw} and \eqref{eq:Gs}, respectively. In Eq.~\eqref{eq:tt}, the correlations functions, $\chi^{\rm ws}_{i j}(q)$ and $\chi^{\rm sw}_{i j}(q)$ are given by
\begin{align}
&\chi^{\rm ws}_{i j}(q) = \frac{1}{2}\sum _k {\rm tr}_4
\left[
G^{\rm sw}(k)\Lambda^{\rm w}_i G^{\rm ws}(k-q)\Lambda^{\rm s}_j 
\right], \label{eq:chiws} 
\end{align}
and $\chi^{\rm sw}_{ij}(q) = \frac{1}{2} \sum _k {\rm tr}_4
[
G^{\rm ws}(k)\Lambda^{\rm s}_i G^{\rm sw}(k-q)\Lambda^{\rm w}_j 
]=\chi^{\rm ws}_{\phi i}(-q)$, 
These correlation functions describe the strength of the direct coupling between fluctuations in the semiconducting wire and superconductor. In Fig.~\ref{fig:diagram2}(a), we show the diagrammatic representations of these correlation functions. 

\begin{figure}[t!]
\includegraphics[width=\columnwidth]{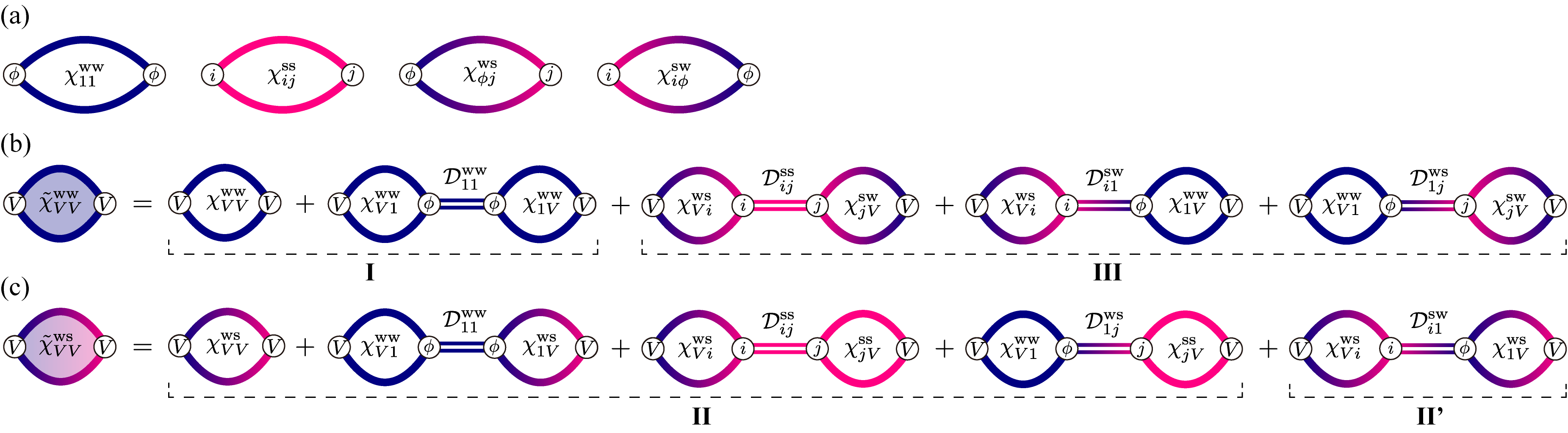}
\caption{(a) Diagrammatic representations of the correlation functions, $\chi^{\rm ww}_{\phi\phi}$, $\chi^{\rm ss}_{ij}$, $\chi^{\rm ws}_{\phi j}$, and $\chi^{\rm sw}_{i\phi}$. (b,c) Renormalized density-density correlation functions, $\tilde{\chi}^{\rm ww}_{VV}$ and $\tilde{\chi}^{\rm ws}_{VV}$, defined in Eq.~\eqref{eq:chi_final}, where the former (latter) function represents the density response of the semiconducting wire to an external perturbation that initially drives the wire (the superconductor). The renormalized functions consist of the bare correlation functions and the fluctuation propagators in the semiconductor and the superconductor, , $\mathcal{D}^{\rm ww}_{11}$, $\mathcal{D}^{\rm ws}_{1 j}$, $\mathcal{D}^{\rm sw}_{i1}$, and $\mathcal{D}^{\rm ss}_{ij}$, where the repeated Roman indices ($i,j$) in the diagrams imply the sum over $1,2,3$. The encircled symbol ``$\phi$'' (``$i,j$'') represents the vertex functions, $\Lambda^{\rm w}_1 = ie\tau_z$ ($\Lambda^{\rm s}_{1,2,3}$), and the vertex $V$ is obtained from $\Lambda^{\rm w}=ie\tau_z$ by replacing $ie$ with $V^{\rm w}$ and $V^{\rm s}$. In (b,c), the labels ``I'', ``II'', and ``III'' denote three processes that influence the density response of the semiconducting wire shown in Fig.~5(b) in the main text. The diagram indicated by ``II$^{\prime}$'' corresponds to the higher order tunneling process of II. The diagrams for the parent superconductor, $\tilde{\chi}^{\rm ss}_{VV}$ and $\tilde{\chi}^{\rm sw}_{VV}$, can be represented in a similar manner with (b) and (c).}
\label{fig:diagram2}  
\end{figure}

Let us introduce the fluctuation propagators, $\mathcal{D}_{ij}^{ab}(q)$, as the matrix form in the basis of $[\delta\phi^{\rm w},\delta\Delta^{{\rm s}+},\delta\Delta^{{\rm s}-},\delta\phi^{\rm s}]$,
\beq
\begin{pmatrix}
\mathcal{D}^{\rm ww}_{11}(q) & \mathcal{D}^{\rm ws}_{11}(q) & 
\mathcal{D}^{\rm ws}_{12}(q) & \mathcal{D}^{\rm ws}_{13}(q) \\
\mathcal{D}^{\rm sw}_{11}(q) & \mathcal{D}^{\rm ss}_{11}(q) & 
\mathcal{D}^{\rm ss}_{12}(q) & \mathcal{D}^{\rm ss}_{13}(q) \\
\mathcal{D}^{\rm sw}_{21}(q) & \mathcal{D}^{\rm ss}_{21}(q) &
\mathcal{D}^{\rm ss}_{22}(q) & \mathcal{D}^{\rm ss}_{23}(q) \\
\mathcal{D}^{\rm sw}_{31}(q) & \mathcal{D}^{\rm ss}_{31}(q) &
\mathcal{D}^{\rm ss}_{32}(q) & \mathcal{D}^{\rm ss}_{33}(q)
\end{pmatrix}
\equiv \begin{pmatrix}
\displaystyle{\frac{e^2}{U_q^{\rm w}} + \chi^{\rm ww}_{11}} & 
\chi^{\rm ws}_{11} & \chi^{\rm ws}_{12} & \chi^{\rm ws}_{13} \\
\chi^{\rm sw}_{11} & 
\displaystyle{\frac{1}{2V_{\rm s}} + \chi^{\rm ss}_{11}} & \chi^{\rm ss}_{12} & \chi^{\rm ss}_{13} \\
\chi^{\rm sw}_{21} & \chi^{\rm ss}_{21} & 
\displaystyle{- \frac{1}{2V_{\rm s}} + \chi^{\rm ss}_{22}} & 
\chi^{\rm ss}_{23} \\
\chi^{\rm sw}_{31} &  \chi^{\rm ss}_{31} &  
\chi^{\rm ss}_{32} & \displaystyle{\frac{e^2}{U_q^{\rm s}} + \chi^{\rm ss}_{33}}
\end{pmatrix}^{-1}.
\label{eq:D2}
\eeq
The matrix elements, $\mathcal{D}^{\rm ww}_{11}$ and $\mathcal{D}^{\rm ss}_{ij}$ ($i,j=1,2,3$), describe the propagators of the bosonic fluctuations in the semiconducting wire and the superconductor, respectively, while the others ($\mathcal{D}^{\rm sw}_{i1}$ and $\mathcal{D}^{\rm ws}_{1j}$) propagate across two subsystems. When all the off-diagonal components are negligible, the diagonal component, $\mathcal{D}^{\rm ww}_{11}$, is the plasmon propagator in the semiconducting wire, while $\mathcal{D}^{\rm ss}_{11}$, $\mathcal{D}^{\rm ss}_{22}$, and $\mathcal{D}^{\rm ss}_{33}$ describe the propagation of the Higgs boson (the amplitude mode), the Nambu-Goldstone boson (the phase mode), and the plasmon, respectively. By substituting Eq.~\eqref{eq:tt} into Eq.~\eqref{eq:action_fluc} and using the expression of the bosonic propagators, the fluctuation action is recast into
\begin{align}
\mathcal{S}_{\rm fluc} = \frac{1}{2}\sum_q\left[
\delta\Sigma^a_i(q)\mathcal{D}^{ab-1}_{ij}(q)\delta\Sigma^b_j(-q)
+\delta\Sigma^{a}_i(q) \chi^{ab}_{iV}(q) V^{b}(-q) 
+V^a(q) \chi^{ab}_{Vj}(q) \delta\Sigma^{b}_j(-q)
+V^{a}(q) \chi^{ab}_{VV}(q) V^{b}(-q)
\right],
\label{eq:sfluct2}
\end{align}
where the repeated indices, $a,b$ and $i,j$, imply the sum over $({\rm s},{\rm w})$ and $1,2,3$, respectively.
The effective action that describes the dynamical response to external perturbation can be derived by integrating out the self-energy fluctuations in Eq.\eqref{eq:sfluct2} as
\begin{align}
\mathcal{S}_{\rm fluc} = \frac{1}{2}\sum_qV^{a}(q)
\left[
\chi^{ab}_{VV}(q) 
- \chi^{ac}_{Vi}\mathcal{D}^{cd}_{ij}(q)\chi^{db}_{jV}(q)
\right]
V^b(-q).
\label{eq:sfluct3}
\end{align}
To this end, the renormalized density-density response functions are obtained as 
\beq
\tilde{\chi}^{ab}_{VV}(q) 
=\frac{\delta^2 \mathcal{S}_{\rm fluc}[V^{\rm w},V^{\rm s}]}{\delta V^a\delta V^b}
= \chi^{ab}_{VV}(q) 
- \chi^{ac}_{Vi}\mathcal{D}^{cd}_{ij}(q)\chi^{db}_{jV}(q).
\label{eq:chi_final}
\eeq
The diagonal components of the tensor, $\tilde{\chi}^{{\rm ww}}_{VV}(q)$ and $\tilde{\chi}^{{\rm ss}}_{VV}(q)$, represent the density response of the semiconducting wire and the superconductor to the density perturbation applied to each subsystem, respectively, while the off-diagonal component, $\tilde{\chi}^{\rm ws}_{VV}$ ($\tilde{\chi}^{\rm sw}_{VV}$), describes the response of the semiconducting wire (the superconductor) to the perturbation applied to the superconductor (the semiconducting wire). The first term in Eq.~\eqref{eq:chi_final} corresponds to the contribution of the quasiparticle excitations to the density response, and the second term reflects the density response mediated by the collective excitations, where $\chi^{ac}_{Vi}$ and $\chi^{db}_{jV}(q)$ represent the direct coupling of external perturbation to the collective modes. These functions, $\chi_{VV}^{ab}$, $\chi_{V i}^{ab}$, and $\chi^{ab}_{jV}$, are obtained from Eqs.~\eqref{eq:chiww}-\eqref{eq:chiws} by replacing the vertex function $\Lambda^{a}_j$ with $\Lambda^{\rm a}\equiv V^{a}\tau_z$.

The diagrammatic representations of $\tilde{\chi}^{\rm ww}_{VV}$ and $\tilde{\chi}^{\rm ws}_{VV}$ are displayed in Figs.~\ref{fig:diagram2}(b) and \ref{fig:diagram2}(c). Their diagrams consist of the bare correlation functions and the fluctuation propagators in the semiconductor and the superconductor, $\mathcal{D}^{\rm ww}_{11}$, $\mathcal{D}^{\rm ws}_{1 j}$, $\mathcal{D}^{\rm sw}_{i1}$, and $\mathcal{D}^{\rm ss}_{ij}$ ($i,j=\pm,\phi$), which are defined in Eq.~\eqref{eq:D2}. The encircled symbol ``$\phi$'' (``$i,j$'') represents the vertex functions, $\Lambda^{\rm w}_1 = ie\tau_z$ ($\Lambda^{\rm s}_{1,2,3}$), and the vertex $V$ is obtained from $\Lambda^{\rm w}=V^{\rm w}\tau_z$ by replacing $ie$ with $V^{\rm w}$ and $V^{\rm s}$. The diagrams labeled by ``I'', ``II'', and ``III'' correspond to three processes that influence the density response of the semiconducting wire shown in Fig.~5(b) in the main text. The diagram ``II$^{\prime}$'' corresponds to the higher order tunneling process of II. The diagrams of the renormalized density response functions for the parent superconductor, $\tilde{\chi}^{\rm ss}_{VV}$ and $\tilde{\chi}^{\rm sw}_{VV}$, can be represented in a similar manner with (b) and (c).

\subsection{Density response functions and electronic Raman spectroscopy}

As mentioned in the main text, the effective dynamical density fluctuations can be measured by electronic Raman spectroscopy. In fact, the generalized structure function of the wire, $S^{\rm w}$, in Raman experiments is related to the imaginary part of the Raman response function $\tilde{\chi}^{ab}_{\gamma\gamma}$~\cite{klein84,mon90,dev07},
\begin{align}
S^{\rm w}({\bm q},\omega) = -\frac{1}{\pi}\left[
1+f_{\rm B}(\omega)
\right]
{\rm Im} \left[ \tilde{\chi}^{{\rm ww}}_{\gamma\gamma}({\bm q},\omega)
+\tilde{\chi}^{{\rm ws}}_{\gamma\gamma}({\bm q},\omega)
\right],
\label{eq:Sw}
\end{align}
where $f_{\rm B}(\omega) = 1/(e^{\omega/T}-1)$ is the bosonic distribution function at temperature $T$. In the functional integral formalism, the Raman response can be incorporated by introducing a source term $V_{\rm R}$ coupled to the Raman density operator, $\hat{\rho}_{\rm R}({\bm q}) \equiv \sum_{{\bm k},\sigma}\gamma_{\bm k}c^{\dag}_{{\bm k}+{\bm q},\sigma}c_{{\bm k},\sigma}$. The Raman vertex function, $\gamma_{\bm k}$ does not depend on $q$ for $q\ll k_{\rm F}$, while it is sensitive to the polarization directions of the incident and scattered light. The effective action for the Raman response is obtained from Eq.~\eqref{eq:sfluct3} by replacing $V^a$ to $V_{\rm R}^a$. Similarly with Eq.~\eqref{eq:chi_final}, the Raman response functions are then given by 
\beq
\tilde{\chi}^{ab}_{\gamma\gamma}(q) 
=\frac{\delta^2 \mathcal{S}_{\rm fluc}[V^{\rm w}_{\rm R},V^{\rm s}_{\rm R}]}{\delta V^a_{\rm R}\delta V^b_{\rm R}}
= \chi^{ab}_{\gamma\gamma}(q) 
- \chi^{ac}_{\gamma i}\mathcal{D}^{cd}_{ij}(q)\chi^{db}_{j \gamma}(q).
\label{eq:chiR}
\eeq
The bare functions, $\chi_{\gamma\gamma}^{ab}$, $\chi_{\gamma i}^{ab}$, and $\chi^{ab}_{j\gamma}$, are obtained from Eqs.~\eqref{eq:chiww}-\eqref{eq:chiws} by replacing the vertex function $\Lambda^{a}_j$ with $\Lambda^{\rm a}\equiv V^{a}_{\rm R}\gamma_{\bm k}\tau_z$. In this work, we consider a parabolic electron band. In this situation, the Raman vertex, $\gamma$, is independent of $k$.

Let us first ignore the cross response functions, $\chi^{\rm ws}(q) = \chi^{\rm sw}(q)=0$, and focus solely on the response of the semiconducting wire with a proximity-induced $\Delta$. This corresponds to the diagrams ``I'' in Fig.~\ref{fig:diagram2}(b). The response function is obtained from Eq.~\eqref{eq:chiR} as 
\beq
\tilde{\chi}^{\rm ww}_{\gamma\gamma}(q) = \chi^{\rm ww}_{\gamma\gamma}(q) 
- \chi^{\rm ww}_{\gamma 1}\mathcal{D}^{\rm ww}_{11}(q)\chi^{\rm ww}_{1\gamma}(q)
\label{eq:chiww_final}
\eeq
Let $\chi_{\rho\rho}$ be the bare density-density response function defined in the main text. In our notation, the bare response functions in Eq.~\eqref{eq:chiww_final} are expressed in terms of the bare density-density response function as $\chi^{\rm ww}_{11}(q)=-e^2\chi_{\rho\rho}(q)$, $\chi^{\rm ww}_{1\gamma}(q) = e\gamma \chi_{\rho\rho}(q)$, and $\chi^{\rm ww}_{\gamma\gamma}(q) = -\gamma^2 \chi_{\rho\rho}(q)$. Then, Eq.~\eqref{eq:chiww_final} reduces to the density-density response function $\tilde{\chi}_{\rho\rho}$
\beq
\tilde{\chi}^{\rm ww}_{\gamma\gamma}(q) = \gamma^2 \frac{\chi_{\rho\rho}(q)}{1-U_q\chi_{\rho\rho}(q)}
= \gamma^2 \tilde{\chi}_{\rho\rho}(q).
\eeq
This is equivalent to the renormalized response function obtained from the Rashba Hamiltonian in Eq.~\eqref{eq:Hbdg} within the random phase approximation. We discuss the main features of the renormalized response function in the Rashba superconductor, $\tilde{\chi}^{\rm ww}_{\gamma\gamma}(q)=\gamma^2\tilde{\chi}_{\rho\rho}(q)$, in the main text and present some supplementary information in Sec.~S2.

\subsection{Collective modes and density response in $s$-wave superconductors}
\label{sec:sc_cm}

Following the previous subsection, let us keep assuming $\chi^{\rm sw}=\chi^{\rm ws}=0$ and focus on the response of the parent superconductor. The renormalized Raman response functions are obtained from Eq.~\eqref{eq:chi_final} as
\begin{gather}
\tilde{\chi}^{\rm ss}_{\gamma\gamma}(q)
= \chi^{\rm ss}_{\gamma\gamma}(q) 
- \chi^{{\rm ss}}_{\gamma i}\mathcal{D}^{\rm ss}_{ij}(q)\chi^{{\rm ss}}_{j\gamma}(q).
\label{eq:chiss}
\end{gather}
The first term describes the excitations of quasiparticles, and the second term contains the propagators of the superconducting order fluctuations and charge density fluctuation,
$\mathcal{D}^{\rm ss}_{ij}(q)$.
In the long wavelength limit $v^{\rm s}_{\rm F}q\ll \Delta^{\rm s}_0$, where $v^{\rm s}_{\rm F}$ is the Fermi velocity of electrons in the parent superconductor, the propagators in the superconducting sector, which are the $3\times 3$ matrix from the lower block in Eq.~\eqref{eq:D2}, reduces to 
\beq
\begin{pmatrix}
\mathcal{D}^{\rm ss}_{11}(q) & 
\mathcal{D}^{\rm ss}_{12}(q) & \mathcal{D}^{\rm ss}_{13}(q) \\
\mathcal{D}^{\rm ss}_{21}(q) &
\mathcal{D}^{\rm ss}_{22}(q) & \mathcal{D}^{\rm ss}_{23}(q) \\
\mathcal{D}^{\rm ss}_{31}(q) &
\mathcal{D}^{\rm ss}_{32}(q) & \mathcal{D}^{\rm ss}_{33}(q)
\end{pmatrix}
\approx \begin{pmatrix}
\displaystyle{\frac{1}{2V^{\rm s}} + \chi^{\rm ss}_{11}} & 0 & 0 \\
0 & 
\displaystyle{- \frac{1}{2V^{\rm s}} + \chi^{\rm ss}_{22}} & 
\chi^{\rm ss}_{23} \\
0 &  \chi^{\rm ss}_{32} & \displaystyle{\frac{e^2}{U_q^{\rm s}} + \chi^{\rm ss}_{33}}
\end{pmatrix}^{-1}.
\eeq
In the long wavelength limit, only the phase and charge sectors can be intrinsically coupled, and the direct couplings between the Higgs and the other sectors are negligible, {\rm e.g.,} $\chi^{\rm ss}_{13}=O(\Delta^{\rm s}_0/\varepsilon^{\rm s}_{\rm F})$~\cite{cea16}. 

The propagator, $\mathcal{D}^{\rm ss}_{11}=[1/2V^{\rm s}+ \chi^{\rm ss}_{11}]$, is the propagator of the Higgs mode that represents the amplitude oscillation of the superconducting gap. The propagator has a pole at the Higgs mode excitation, $\omega = \sqrt{(2\Delta^{\rm s}_0)^2+(vq)^2}$, where the velocity is given by $v=v^{\rm s}_{\rm F}/\sqrt{d}$ in $d$ spatial dimension. 
The propagator of the phase mode, $[- \frac{1}{2V^{\rm s}} + \chi^{\rm ss}_{22}]^{-1}$, has a pole at $\omega = vq$.
For $\omega\sim \Delta^{\rm s}_0$ and $\Delta^{\rm s}_0\ll \varepsilon^{\rm s}_{\rm F}$, the propagator of the Coulomb potential, $[\frac{1}{U_q^{\rm s}} + \chi^{\rm ss}_{33}]^{-1}$, the propagator is recast into 
\beq
 \left[\frac{e^2}{U_{q}^{\rm s}}+ \chi^{\rm ss}_{33}(q)\right]^{-1}
\approx\left[ \frac{q^2}{4\pi}+2e^2N^{\rm s}_{\rm F}\lambda(\omega)\right]^{-1}.
\eeq
where $N^{\rm s}_{\rm F}$ is the normal state density of states in the parent superconductor and the Tsuneto function, $\lambda(\omega)$, is defined as
\beq
\lambda(\omega) = \left(\Delta^{{\rm s}}_0\right)^2 \int^{\infty}_{\Delta_0} \frac{d\varepsilon}{\sqrt{\varepsilon^2-(\Delta^{\rm s}_0)^2}}
\frac{\tanh(\varepsilon/2T)}{\varepsilon^2-\omega^2/4}
\eeq
The function sharply peaks at $\omega = 2\Delta^{\rm s}_0$ for ${\bm q}\approx {\bm 0}$. In the high frequency region, $\omega \gg \Delta^{\rm s}_0$, the propagator has a pole at the plasma frequency, $\omega = \omega_{\rm p}$.
We note that introducing the coupling $\chi^{\rm ss}_{23}$ and $\chi^{\rm ss}_{32}$ gaps out the phase mode to the plasma frequency, $\omega_{\rm p}=\sqrt{4\pi e^2 n_0/m_{\rm s}}$, through the Anderson-Higgs mechanism, where $n_0$ is the electron density. Hence, only the low-lying collective excitation in the parent superconductor is the Higgs mode. 

\begin{figure}[t!]
\includegraphics[width=\columnwidth]{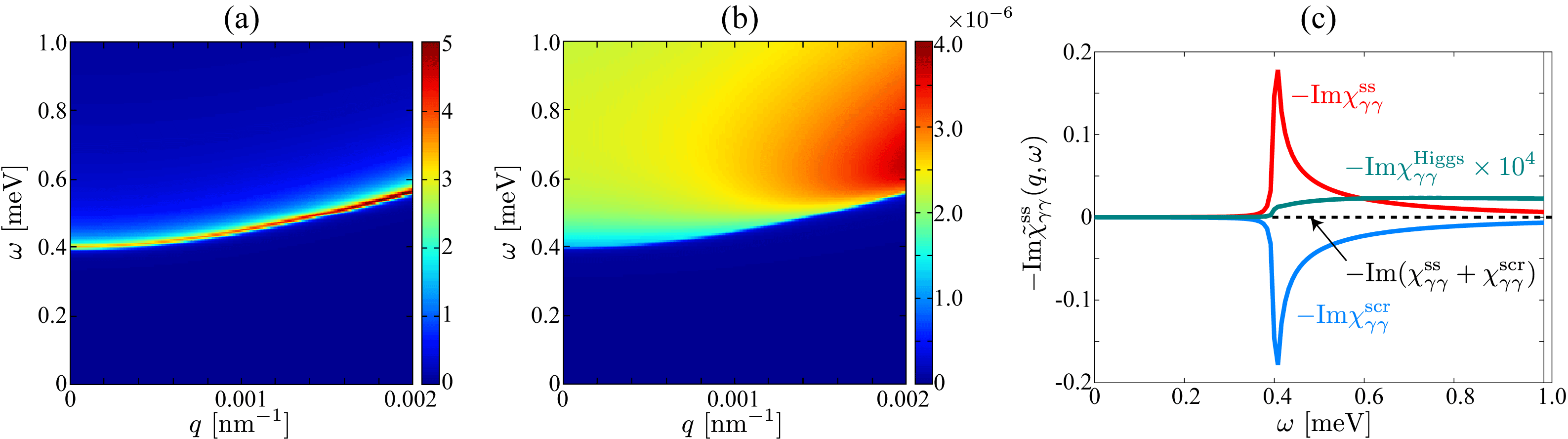}
\caption{(a) Spectral function of the superconducting gap amplitude mode (the Higgs mode), $-{\rm Im}\mathcal{D}^{\rm ss}_{11}(q,\omega)$, in the parent superconductor. (b) Renormalized density-density response function in Eq.~\eqref{eq:chiss}, $-{\rm Im}\tilde{\chi}^{\rm ss}_{\gamma\gamma}(q,\omega)$. In (c), we plot the bare density response function ($-{\rm Im}\chi^{\rm ss}_{\gamma\gamma}$), the screening effect ($-{\rm Im}\chi^{\rm scr}_{\gamma\gamma}$), and the Higgs contribution to the density response ($-{\rm Im}\chi^{\rm Higgs}_{\gamma\gamma}$) at $q=1.0\times 10^{-4}~[{\rm nm}^{-1}]$. The dashed line corresponds to the sum of $\chi^{\rm ss}_{\gamma\gamma}$ and $\chi^{\rm scr}_{\gamma\gamma}$, which stays zero within numerical accuracy.}
\label{fig:higgs}  
\end{figure}

In Fig.~\ref{fig:higgs}, we show the collective excitation spectrum and density response of the parent superconductor. Here we set the parameters to be $\Delta^{\rm s}_0=0.2~{\rm meV}$, $\mu^{\rm s}=1~{\rm eV}$, and $m_{\rm s}=m_{\rm e}$. As we are interested in the vicinity of the TPT, we ignore the depairing effect due to an applied magnetic field. It is seen from Fig.~\ref{fig:higgs}(a) that the spectral function of the renormalized propagator, $-{\rm Im}\mathcal{D}_{11}$, exhibits sharp dispersion starting from $\omega = 2\Delta^{\rm s}_0$, indicating the existence of the long-lived gap-amplitude oscillation mode. Figure~\ref{fig:higgs}(b) shows the renormalized density response function, $-{\rm Im}\tilde{\chi}^{\rm ss}_{\gamma\gamma}(q,\omega)$, defined in Eq.~\eqref{eq:chiss}. To understand the density response, let us decompose the response function in Eq.~\eqref{eq:chiss} as 
\beq
\tilde{\chi}^{\rm ss}_{\gamma\gamma}(q,\omega)
= {\chi}^{\rm ss}_{\gamma\gamma}(q,\omega) + 
{\chi}^{\rm scr}_{\gamma\gamma}(q,\omega) + 
{\chi}^{\rm Higgs}_{\gamma\gamma}(q,\omega).
\label{eq:chiss2}
\eeq
The bare response function in the first term ($\chi^{\rm ss}_{\gamma\gamma}$) describes the pair excitations. The second term represents the screening effect of the long-range Coulomb interaction,
\begin{align}
{\chi}^{\rm scr}_{\gamma\gamma}(q,\omega) = &
\chi^{\rm ss}_{\gamma 2}(q,\omega)\mathcal{D}^{\rm ss}_{22}(q,\omega)\chi^{\rm ss}_{2\gamma}(q,\omega) 
+ \chi^{\rm ss}_{\gamma 2}(q,\omega)\mathcal{D}^{\rm ss}_{23}(q,\omega)\chi^{\rm ss}_{3\gamma}(q,\omega) \nn \\
&+ \chi^{\rm ss}_{\gamma 3}(q,\omega)\mathcal{D}^{\rm ss}_{32}(q,\omega)\chi^{\rm ss}_{2\gamma}(q,\omega) 
+ \chi^{\rm ss}_{\gamma 3}(q,\omega)\mathcal{D}^{\rm ss}_{33}(q,\omega)\chi^{\rm ss}_{3\gamma}(q,\omega) ,
\end{align}
which are mediated by the propagators of the phase and charge density fluctuations. The last term in Eq.~\eqref{eq:chiss2} is the contribution from the Higgs excitation,
\beq
{\chi}^{\rm Higgs}_{\gamma\gamma}(q,\omega) = \chi^{\rm ss}_{\gamma 1}(q,\omega)\mathcal{D}^{\rm ss}_{11}(q,\omega)\chi^{\rm ss}_{1\gamma}(q,\omega) .
\label{eq:chi_higgs}
\eeq
We plot the three terms of Eq.~\eqref{eq:chiss2} in Fig.~\ref{fig:higgs}(c), where we fix $q=1.0\times 10^{-4}~[{\rm nm}^{-1}]$. The bare response function ($\chi^{\rm ss}_{\gamma\gamma}$) has a sharp peak at $\omega = 2\Delta^{\rm s}_0$ corresponding to the threshold energy of the pair excitations. It is seen from Fig.~\ref{fig:higgs}(c) that the density response due to the pair excitations is fully screened by the superconducting phase and charge density responses, $\chi^{\rm scr}_{\gamma\gamma}$. Only the tiny but nonvanishing density response is mediated by the Higgs excitation. The coupling of the Higgs excitation to an external density perturbation is described by $\chi^{\rm ss}_{13}$, which reduces to 
\beq
\chi^{\rm ss}_{13}(\omega) \approx \sum_{\bm k} \frac{\xi^{\rm s}_k}{2E^{\rm s}_{k}}\frac{\tanh(E^{\rm s}_k/2T)}{(\omega + i0_+)^2-(2E^{\rm s}_k)^2},
\eeq
in the long wavelength limit $q\ll \Delta^{\rm s}_0/v^{\rm s}_{\rm F}$, where $0_+$ is an infinitesimal constant. As the integrand is an odd function on $\xi^{\rm s}_k$, the correlation function vanishes in the limit of $\Delta^{\rm s}_0\ll \varepsilon_{\rm F}^{\rm s}$, where the normal-state density of state approximately exhibits the particle-hole symmetry. Therefore, the coupling of the Higgs mode and external perturbation originates from the approximate particle-hole symmetry breaking with an order of $O(\Delta^{\rm s}_0/\varepsilon_{\rm F}^{\rm s})$ and the Higgs contribution in Eq.~\eqref{eq:chi_higgs} is negligibly small.

\subsection{Effect of pair tunneling between semiconductor and superconductor}

Lastly, we examine the pair tunneling effect in semiconductor-superconductor hybrid systems by solving the full renormalized response functions in Eq.~\eqref{eq:chi_final}. There are two contributions that describe the dynamical density response of the semiconducting wire,
\begin{align}
\tilde{\chi}^{\rm ww}_{\gamma\gamma}(q) 
=& \chi^{\rm ww}_{\gamma\gamma}(q) 
- \chi^{\rm ww}_{\gamma 1}\mathcal{D}^{\rm ww}_{11}(q)\chi^{\rm ww}_{1\gamma}(q)
- \chi^{\rm ws}_{\gamma i}\mathcal{D}^{\rm ss}_{ij}(q)\chi^{\rm sw}_{j\gamma}(q)
- \chi^{\rm ws}_{\gamma i}\mathcal{D}^{\rm sw}_{i1}(q)\chi^{\rm ww}_{1\gamma}(q)
- \chi^{\rm ww}_{\gamma 1}\mathcal{D}^{\rm ws}_{1j}(q)\chi^{\rm sw}_{j\gamma}(q),
\label{eq:chiwwVV} \\
\tilde{\chi}^{\rm ws}_{\gamma\gamma}(q) 
=& \chi^{\rm ws}_{\gamma\gamma}(q) 
- \chi^{\rm ww}_{\gamma 1}\mathcal{D}^{\rm ww}_{11}(q)\chi^{\rm ws}_{1\gamma}(q)
- \chi^{\rm ws}_{\gamma i}\mathcal{D}^{\rm ss}_{ij}(q)\chi^{\rm ss}_{j\gamma}(q)
- \chi^{\rm ww}_{\gamma 1}\mathcal{D}^{\rm ws}_{1j}(q)\chi^{\rm ss}_{j\gamma}(q)
- \chi^{\rm ws}_{\gamma i}\mathcal{D}^{\rm sw}_{i1}(q)\chi^{\rm ws}_{1\gamma}(q).
\label{eq:chiwwVV2}
\end{align}
where the repeated indices $i,j$ imply the sum over $1,2,3$. Equations~\eqref{eq:chiwwVV} and \eqref{eq:chiwwVV2} describe the response of the semiconducting wire when a perturbation is applied to the semiconducting wire and the parent superconductor, respectively, and their diagrammatic representations are shown in Fig.~\ref{fig:diagram2}(b) and \ref{fig:diagram2}(c). As mentioned in Fig.~\ref{fig:diagram2} and Fig.~5(a) in the main text, the normalized dynamical density response functions can be decomposed into three terms as
\begin{align}
&\tilde{\chi}^{\rm I}_{\gamma\gamma}(q) 
= \chi^{\rm ww}_{\gamma\gamma}(q) 
- \chi^{\rm ww}_{\gamma 1}\mathcal{D}^{\rm ww}_{11}(q)\chi^{\rm ww}_{1\gamma}(q),\\
&\tilde{\chi}^{\rm II}_{\gamma\gamma}(q) 
= \tilde{\chi}^{\rm ws}_{\gamma\gamma}(q) , \\
&\tilde{\chi}^{\rm III}_{\gamma\gamma}(q) 
= -\chi^{\rm ws}_{\gamma i}\mathcal{D}^{\rm ss}_{ij}(q)\chi^{\rm sw}_{j\gamma}(q)
- \chi^{\rm ws}_{\gamma i}\mathcal{D}^{\rm sw}_{i1}(q)\chi^{\rm ww}_{1\gamma}(q)
- \chi^{\rm ww}_{\gamma 1}\mathcal{D}^{\rm ws}_{1j}(q)\chi^{\rm sw}_{j\gamma}(q).
\end{align}

\begin{figure}[t!]
\includegraphics[width=\columnwidth]{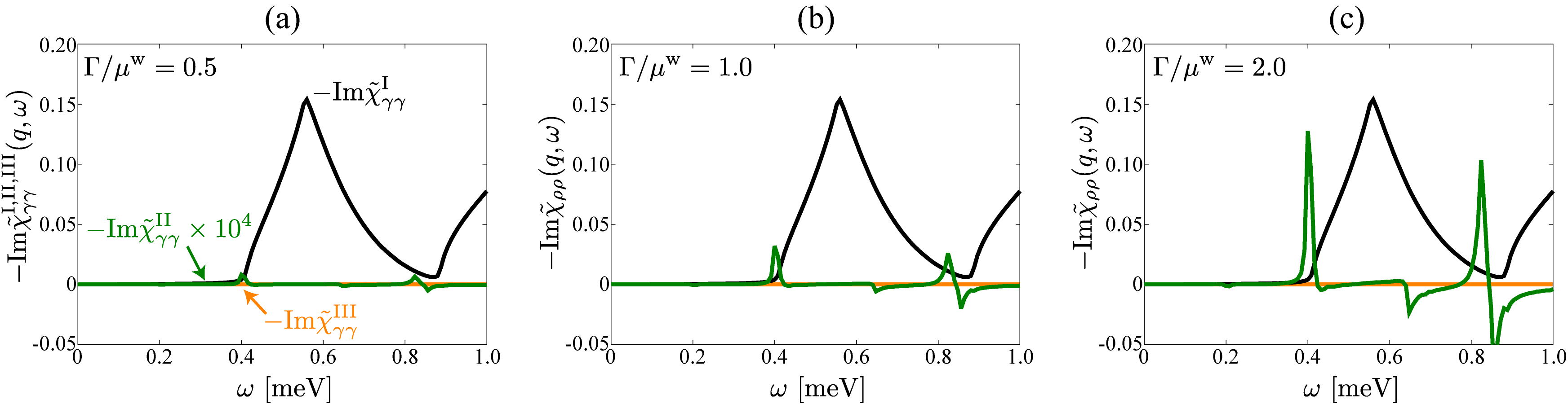}
\caption{Tunneling energy dependences of renormalized Raman functions, $\tilde{\chi}^{\rm I}_{\gamma\gamma}=\tilde{\chi}_{\rho\rho}$, $\tilde{\chi}^{\rm II}_{\gamma\gamma}$, and $\tilde{\chi}^{\rm III}_{\gamma\gamma}$, in the semiconducting wire at $\delta B = -0.18~{\rm T}$: (a) $\Gamma/\mu^{\rm w}=0.5$, (b) $1.0$, and (c) $2.0$. Three curves denoted by $\tilde{\chi}^{\rm I}_{\gamma\gamma}$, $\tilde{\chi}^{\rm II}_{\gamma\gamma}$, and $\tilde{\chi}^{\rm III}_{\gamma\gamma}$ correspond to the diagrams I, II, and III shown in Figs.~\ref{fig:diagram2}, respectively. Here we set $\gamma = 1$.}
\label{fig:density}  
\end{figure}

Figures~\ref{fig:density} show the Raman response functions in the semiconducting wire for different tunneling energies, $\Gamma/\mu^{\rm s}=0.5$, $1.0$, and $2.0$. Here, we take the same parameters as those in Secs.~S3.1 and S3.2. We choose $m_{\rm eff}=0.014m_{\rm e}$, $\alpha = 0.5~{\rm eV}$\AA, $g=50$, and $\mu^{\rm s} = 1.0~{\rm meV}$~\cite{lut18} for the semiconducting wire, and $\Delta^{\rm s}_0=0.2~{\rm meV}$, $\mu^{\rm s}=1~{\rm eV}$, and $m_{\rm s}=m_{\rm e}$ for the parent superconductor. Figure~\ref{fig:density} demonstrates that the entire profile of the Raman spectrum, which is the sum of $\tilde{\chi}^{\rm I}_{\gamma\gamma}$, $\tilde{\chi}^{\rm I}_{\gamma\gamma}$, and $\tilde{\chi}^{\rm I}_{\gamma\gamma}$, is insensitive to the value of the tunneling energy $\Gamma$ and the Raman spectrum is dominated by the response function, $-{\rm Im}\tilde{\chi}^{\rm I}_{\gamma\gamma} = - {\rm Im}\chi^{\rm ww}_{\rho\rho}$. The tunneling processes of the Cooper pair fluctuations driven in the parent superconductor, which are denoted by $-{\rm Im}\tilde{\chi}^{\rm II}_{\gamma\gamma}$, exhibit a peak at $\omega=0.4~{\rm meV}$, which corresponds to the resonant frequency to the threshold energy of the pair excitations or the Higgs excitation, as discussed in Sec.~S3.2. The resonant peak of $-{\rm Im}\tilde{\chi}^{\rm II}_{\gamma\gamma}$ at $\omega = 0.4~{\rm meV}$ increases with $\Gamma$, while the contribution remains negligible for $\Gamma\lesssim\mu^{\rm s}$. Therefore, we conclude that the collective mode excitations in the parent superconductor provide little contribution to the density response, and the genralized structure factor measured in electronic Raman scattering experiments reduces to 
\begin{align}
S^{\rm w}(q,\omega) 
\approx -\frac{1}{\pi}\left[
1+f_{\rm B}(\omega)
\right]
{\rm Im} \tilde{\chi}^{{\rm ww}}_{\gamma\gamma}(q,\omega).
\end{align}
Thus, the field evolution of the Raman response function can detect the TPT even when considering the fluctuation effects in the parent superconductor.
\end{widetext}

\bibliography{biblio}

\end{document}